\documentclass[
showpacs,
amsmath,amssymb,
aps,
twocolumn,
prb,
floatfix,
nofootinbib
]{revtex4-2}
\usepackage{xcolor}
\usepackage{lipsum, babel}
\usepackage{graphicx}
\usepackage{amssymb}
\usepackage{dcolumn}
\usepackage{tabularray}
\UseTblrLibrary{booktabs}

\begin{document}

\title{Buzdin,  Shapiro and Chimera Steps in  $\varphi_0$ Josephson Junctions. II. Bifurcation, Switching, and Hysteresis.
 }
\author{M. Nashaat$^{1,2,*}$, E. Kovalenko$^{3}$, and Yu. M. Shukrinov$^{1,4,5,\dagger}$}

\affiliation{$^1$\mbox{BLTP, JINR, Dubna, Moscow region, 141980, Russia}\\
$^2$ \mbox{Department of Physics, Faculty of Science, Cairo University, 12613, Giza, Egypt}\\
$^3$  \mbox{Center for the Development of Digital Technologies, Krasnogorsk, Russia} \\
$^4$ \mbox{Dubna State University, Dubna, Russia}\\	
$^5$  \mbox{Moscow Institute of Physics and Technology, Dolgoprudny 141700, Russia}\\
$^{*}$ majed@sci.cu.edu.eg; $^{\dagger}$ shukrinv@theor.jinr.ru
}

\date{\today}

\begin{abstract}
The dynamics of magnetization and current-voltage characteristics of the superconductor-ferromagnet-superconductor $\varphi_0$ Josephson junction in the presence of external electromagnetic radiation have been studied. Effects of radiation magnetic component are investigated in the frameworks of one- and two-signal models. The implementation of two types of dynamical states of magnetization is demonstrated. These states have a phase shift of $\pi$ in the synchronization region of magnetic precession and Josephson oscillations and differ in the nature of their time dependence.  Transitions between these states with increasing and decreasing bias current show hysteresis, which is reflected in the bifurcation diagram and the current-voltage characteristics. We also provide an experimental way to test the obtained results by measuring the phase shift in voltage temporal dependence at fixed current value for both  sweeping directions. The results obtained can find application in various fields of superconducting spintronics and quantum computing.
\end{abstract}
\maketitle

\section{Introduction}

The coexistence and mutual influence of superconductivity and magnetism is one of the most pressing problems in condensed matter physics \cite{linder15,Eschrig2019,shukrinov22ufn,melnikov22ufn,Golovchanskiy2020,Golovchanskiy2023}. An important achievement in this area is the implementation of the coupling between the superconducting phase and magnetic moment in superconductor-ferromagnet-superconductor Josephson junctions (JJ)  with strong spin-orbit coupling \cite{buzdin08,konschelle09}.  A series of interesting results in this field describing anomalous or $\varphi_{0}$ junctions with the phase shift in the ground state and the Josephson junctions on topological insulators are obtained recently \cite{buzdin08,shukrinov22ufn,Hasan22,Tanaka2009,Linder2010,Golubov2015,Dolcini2015,Zyuzin2016}

Experimentally it easier to break the time-reversal symmetry by a magnetic field applied to the Josephson junction. In this case the junction has  a phase shift caused by the Zeeman effect \cite{Mayer20,Szombati2016,Assouline2019,Murani2017}. However, it is very attractive  to realize the anomalous $\varphi_{0}$ junction in the structures with ferromagnetic layer, and open up many possibilities for applications of such structures for magnetization control \cite{bobkova-rev,rabinovich2019resistive, Nashaat2019,Guarcello2020,Bobkov23}. This leads to a number of different  directions in superconducting spintronics, based on reversing the magnetic moment of the ferromagnetic layer \cite{shukrinov-apl17,Bobkova2020}, phase batteries \cite{pal2019}, diode effect \cite{Trahms2023,Narita22}, cryogenic spintronics devices \cite{Guarcello2020,Guarcello23}. Interesting perspectives are opened by applications based on  Kapitza pendulum features demonstrated by the $\varphi_0$ junction \cite{shukrinov-epl18}, as well as the unique nonlinear phenomena \cite{shukrinov-prb22,shukrinov-bjnano-22}.

Furthermore, ferromagnetic resonance (FMR) manifests the reach physics in the anomalous Josephson junctions. In Ref.\cite{Shukrinov2019re} it is shown that as a current sweep along the $IV$-characteristics of the $\varphi_{0}$ junction, it leads to regular magnetization dynamics with a series of specific phase trajectories related to a direct coupling of the magnetic moment and the Josephson oscillations. The magnetization dynamics in SFS thin film structures by ferromagnetic resonance spectroscopy was investigated in Ref.\cite{Golovchanskiy2023}. The one-dimensional anisotropic action of superconducting torque on magnetization dynamics was experimentally established. The authors results support the recently proposed by M. Silaev the mechanism of the superconducting torque formation via the interplay between the superconducting imaginary conductance and magnetization precession at superconductor-ferromagnet interfaces. So, S-F-S systems provide the playground for Anderson-Higgs mass generation of boson quasiparticles in high-energy Standard Model and in condensed-matter systems \cite{Silaev22,Golovchanskiy2023}.

Recently, it was demonstrated in Ref.\cite{Coraiola23}, the JJ's supercurrent can be nonlocally controlled by the phase difference of another JJ. This occurs when the two JJs sharing a single superconducting electrode and are coherently connected forming the Andreev molecules. The results demonstrate that even in the case of a zero local phase difference, the nonlocal phase control produces a finite supercurrent. Strong magnetic field and ferromagnetic material are not necessary for this realization.

In Ref.\cite{Bobkov23}, the authors demonstrate the possibility of controlling the magnetic states in chain of $\varphi_{0}$ junctions.  The static and dynamic magnetic properties of such a system reveals the  manifestation  of an n-level system, in which the energies of the levels are determined by only projections of the total magnetic moment $\sum M_{i}$ onto the easy magnetic axis. Also, it is shown that the total magnetic moment can be controlled by a superconducting current. Another possibility for influencing the properties of the $\varphi_0$ junction arises under external electromagnetic radiation, leading to synchronization of Josephson oscillations and magnetic precession in the ferromagnetic layer \cite{sara2022}.

Our previous work \cite{shukrinov-prb24} revealed important findings after taking into account the interaction of the magnetic field of microwave radiation with the magnetic moment of the ferromagnetic layer. It leads to a number of unique resonant and synchronization phenomena, in particular, the manifestation of two mechanisms of synchronization of Josephson oscillations and magnetic precession. Due to the coupling of superconductivity and magnetism in this system, the precession of the magnetic moment of the ferromagnetic layer, caused by the magnetic component of the external radiation, can synchronize Josephson oscillations, leading to the appearance of a special type of steps in the current-voltage characteristic, completely different from the known Shapiro steps. These steps were called as Buzdin steps in the case when the system is driven only by the magnetic component of radiation, and chimera (due to their different and composite mechanism of the step formation) when both components are taken into account. When the Josephson or external radiation frequency approaches the ferromagnetic one, then the mutual influence of Josephson and Kittel ferromagnetic resonances  occurs \cite{shukrinov-prb24}.

In this work, the sweeping of bias current and magnetization dynamics along the chimera step in the $IV$-characteristics of $\varphi_0$ junction in the Josephson ferromagnetic resonance region is analyzed. We show two-bubble structure in the magnetization and spikes in the $IV$-characteristics reflecting the corresponding transitions between magnetization bubbles. These transitions demonstrate the creation of two different magnetization states caused by bifurcations in magnetic dynamics of ferromagnetic layer.  The manifestation of a novel type of hysteresis related to the transitions between these states is shown. We demonstrate also a possibility of switching between the found magnetization dynamical states by electric current pulse and propose an experimental testing of the observed phenomena.

The paper is organized as follows. The model and methods are introduced in Section II. In Section III we demonstrate  the bifurcation between magnetization dynamical states and discuss the origin and specific features of two states of magnetization along the chimera step. The switching of states by current pulse along the chimera step for decreasing and increasing bias current are demonstrated in Section IV. In Section V we discuss the hysteretic behaviour of magnetization along the steps and its variation by spin-orbit coupling and radiation parameters. Results for one-signal model are presented in Section VI.  The $\pi$ phase shift in supercurrent and voltage dynamics along the chimera step and possible application of the obtained results  are discussed in Section VII. Finally, we come to the conclusions.

\section{Model and Methods}

\begin{figure}[h!]
	\centering
	\includegraphics[width=1\linewidth]{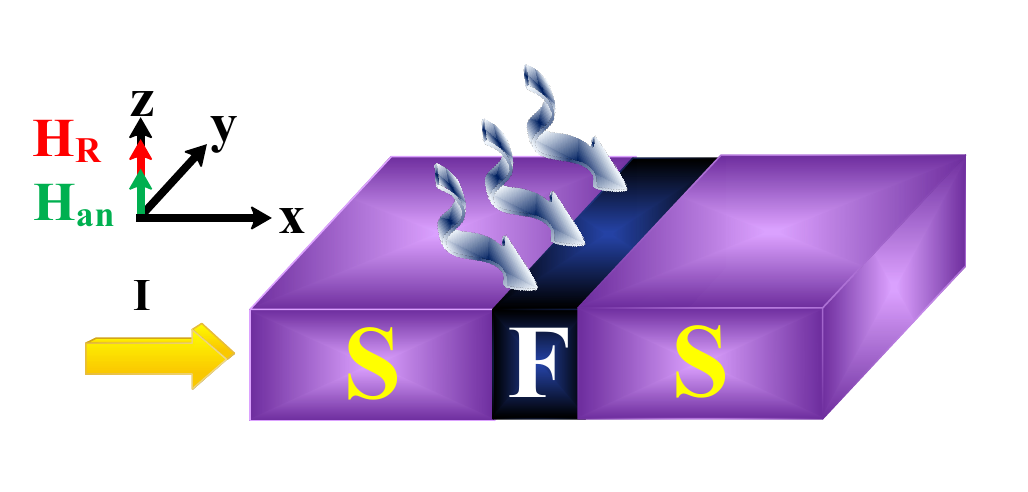}
	\caption{Schematic diagram depicting $\varphi_{0}$ under the influence of external electromagnetic radiation. The magnetic field component ($H_{R}$), and the anisotropic field ($H_{an}$) are in z-direction, while the bias current $I$ is along the x-direction.}
	\label{1}
\end{figure}

In the proposed $\varphi_0$ JJ the easy axis of the ferromagnet and the gradient of the spin-orbit potential are directed along the same axis $z$ (see Fig.\ref{1}). In this case, the phase shift is proportional to the $y$-component of the magnetic moment of the ferromagnet $\varphi_0 = rM_y/M_0$, where $r$ characterizes the magnitude of the spin-orbit interaction \cite{konschelle09}, $M_0 = ||\textbf{M}||$ is the saturation magnetization. The junction is under linearly polarized electromagnetic radiation, which magnetic component $\mathbf{H}_R$ is also parallel to the $z$ axis.

Dynamics of the magnetic moment is determined by the Landau-Lifshitz-Gilbert (LLG) equation \cite{lifshitz1991course}:
\begin{eqnarray}
	\frac{d\mathbf{M}}{dt}&=&-\gamma \mathbf{M}\times \mathbf{H}_{eff}+\frac{\alpha }{M_0}\left( \mathbf{M}\times \frac{d\mathbf{M}}{dt}\right),
	\label{LLG}
\end{eqnarray}
where $\gamma$ is the gyromagnetic ratio, $\alpha$ is the Gilbert damping constant. The effective field is given by \cite{shukrinov-prb24}

\begin{eqnarray}
	\textbf{H}_{eff}&=
	&\frac{K}{M_0}Gr\sin \left( \varphi - r\frac{M_y}{M_0}\right)\hat{\textbf{y}} + \\  \nonumber
	& &+\left( \frac{K}{M_0}\frac{M_z}{M_0} + H_R\sin(\Omega_Rt)\right)\hat{\textbf{z}},
	\label{Heff}
\end{eqnarray}

where ${G=E_J/(K\mathcal{V})}$ is the ratio of the Josephson energy to the magnetic anisotropy energy, $\varphi$ is the phase difference between the superconducting edges of the junction. The effective field includes the interaction between the magnetic moment of a ferromagnet and the magnetic field of external radiation

\begin{equation}
	W_R = - (\mathbf{M},\mathbf{H}_R),
\end{equation}
where $H_R$ is the amplitude of the magnetic component of radiation $\textbf{H}_{R} = (0, 0, H_{R}\sin(\Omega_{R}t))$, $\Omega_{R}$ — radiation frequency.

To describe the coupled dynamics of the magnetization and the superconducting phase difference, we solve the LLG equation together with the RCSJ equation. In this study, we consider a case of the current biased JJ, where the external current $I$ flows through the system according to the extended RCSJ model~\cite{tinkham2004introduction}. In our calculations we change the current direction, decreasing and increasing it in some interested intervals. The total system of equations in the dimensionless form that describes the magnetic and phase dynamics for $\varphi_0$ JJ is given by \cite{shukrinov-prb24}:

\begin{align} \label{LLGJ}
	\begin{split}		
		\dot{m_{x}} =& \frac{1}{\alpha^2 + 1}\{\omega_{F}[-m_{y}m_{z} + Grm_{z}\sin(\varphi-rm_{y})- \\
		-& \alpha(m_{x}m_z^2 + Grm_{x}m_{y}\sin(\varphi-rm_{y}))]- \\
		-& h_{R}(\alpha m_{x}m_{z} + m_{y})sin(\omega_{R}t)\}, \nonumber
	\end{split}\\
	\begin{split}
		\dot{m_{y}} =& \frac{1}{\alpha^2 + 1}\{\omega_{F}[
		m_{x}m_{z}- \\
		-& \alpha(m_{y}m_{z}^2 - Gr(m_{x}^2 + m_{z}^2)sin(\varphi-rm_{y}))]- \\
		-& h_{R}(\alpha m_{y}m_{z} - m_{x})sin(\omega_{R}t)\}, \nonumber
	\end{split}\\
	\begin{split}
		\dot{m_{z}} =& \frac{1}{\alpha^2 + 1}\{\omega_{F}[-Grm_{x}sin(\varphi-rm_{y}) \\
		-& \alpha(
		Grm_{y}m_{z}sin(\varphi-rm_{y}) - m_{z}(m_{x}^2 + m_{y}^2))] \\
		+& h_{R}\alpha (m_{x}^2 + m_{y}^2)sin(\omega_{R}t)\}, \nonumber
	\end{split}\\
	\begin{split}
		\dot{V} =& \left[ I+A\sin(\omega_Rt)-V(t)+r\dot{m}_{y}-\sin (\varphi -rm_{y})\right]/\beta_c, \nonumber
	\end{split}\\
	\begin{split}
		\dot{\varphi} =& V(t),
	\end{split}
\end{align}
where $m_{i}=\frac{M_{i}}{M_0}, (i\equiv x,y,z)$, $\beta_{c}=2eI_{c}CR^{2}/\hbar$ is the McCumber parameter. Here, time is normalized in units $\omega _c^{-1}$, $\omega _c=2eI_cR/\hbar $ is a characteristic frequency of the junction. The ferromagnetic resonance frequency $\Omega _F = K\gamma /M_0$, the frequency of external radiation, and amplitude of magnetic component $\mathbf{H_R}=(0,0,h_R\sin(\Omega_Rt)$  are normalized to $\omega _c$, so that  $\omega _F= \frac{\Omega _F}{\omega _c}$, $\omega _R=\frac{\Omega _R}{\omega _c}$, and $h_R=\frac{\gamma }{\omega _c}H_R$. $A$ is the amplitude of the current caused by the electric component of external radiation, and it is normalized to $I_c$; the external current $I$ is also in the units of $I_c$, and the voltage $V$ in the units of $V_c=I_{c}R$. So, the Josephson frequency $\omega _J=V$, where V denotes the time average of the instantaneous voltage $V(t)$. The last term in the current equation (\ref{LLGJ}) was derived in the framework of microscopic theory for anomalous Josephson junction in Ref. \cite{rabinovich2019resistive}. In our simulations, if it is not mentioned, we use the following model parameters: $G=0.01$, $\alpha=0.01$, $\beta_c=25$, $\omega_F=0.5$, and $r=0.4$.

In the current study we use two schemes of calculations. The first scheme is based on two signal model, in which the electric field of radiation and the magnetic field are taken into account independently. In the second scheme both the electric and magnetic components are related to each other and come from the same radiation signal. As it is demonstrated in Sec. VI, the obtained results for one-signal model exhibit a good agreement with those obtained in the two-signal model.

\section{Two dynamical states of magnetization along the chimera step}

As it was demonstrated in Ref.\cite{shukrinov-prb24}, the magnetization precession is changing along the Buzdin and chimera steps, though the Josephson oscillations are locked to the external electromagnetic radiation. The $m^{max}_y(I)$ dependence demonstrates the specific "bubble-like" feature in the current interval corresponded to the step. Here we show that the variation of $\varphi_0$ junction parameters, in particular, increasing the spin-orbit interactions, or changing direction of bias current, shows more complex behaviour. The $IV$-characteristics along the Buzdin and chimera steps demonstrate the spikes related to the transitions between different dependencies of $m^{max}_y(I)$ with increasing and decreasing the bias current.

Let us first examine the bifurcation $m_{y}^{BF}(I)$  and average of $m_{y}^{av}(I)$ along the chimera step. To realize it, we solve the system of equations (\ref{LLGJ}) numerically. In Fig.2(a) we present the results of one loop calculations (increasing bias current from 0 till 1.2 and back) for the $IV$- characteristic and the maximal value of $m_{y}^{max}(I)$ for the $\varphi_{0}$ junction under external electromagnetic radiation.  As we see, $m_{y}^{max}(I)$ demonstrates split resonance peak in the region of $V=\omega_F=0.5$, its subharmonic and second harmonic. The chimera step which appears at $V=\omega_R=0.485$, has a jump (surge), which coincides in the position with a specific  feature in the $m_{y}^{max}(I)$ dependence (see hollow arrow in the inset). The $m_{y}^{max}(I)$ demonstrates the usual bubble structure along the chimera step.

\begin{figure}[h!]
	\centering
	\includegraphics[width=0.9\linewidth]{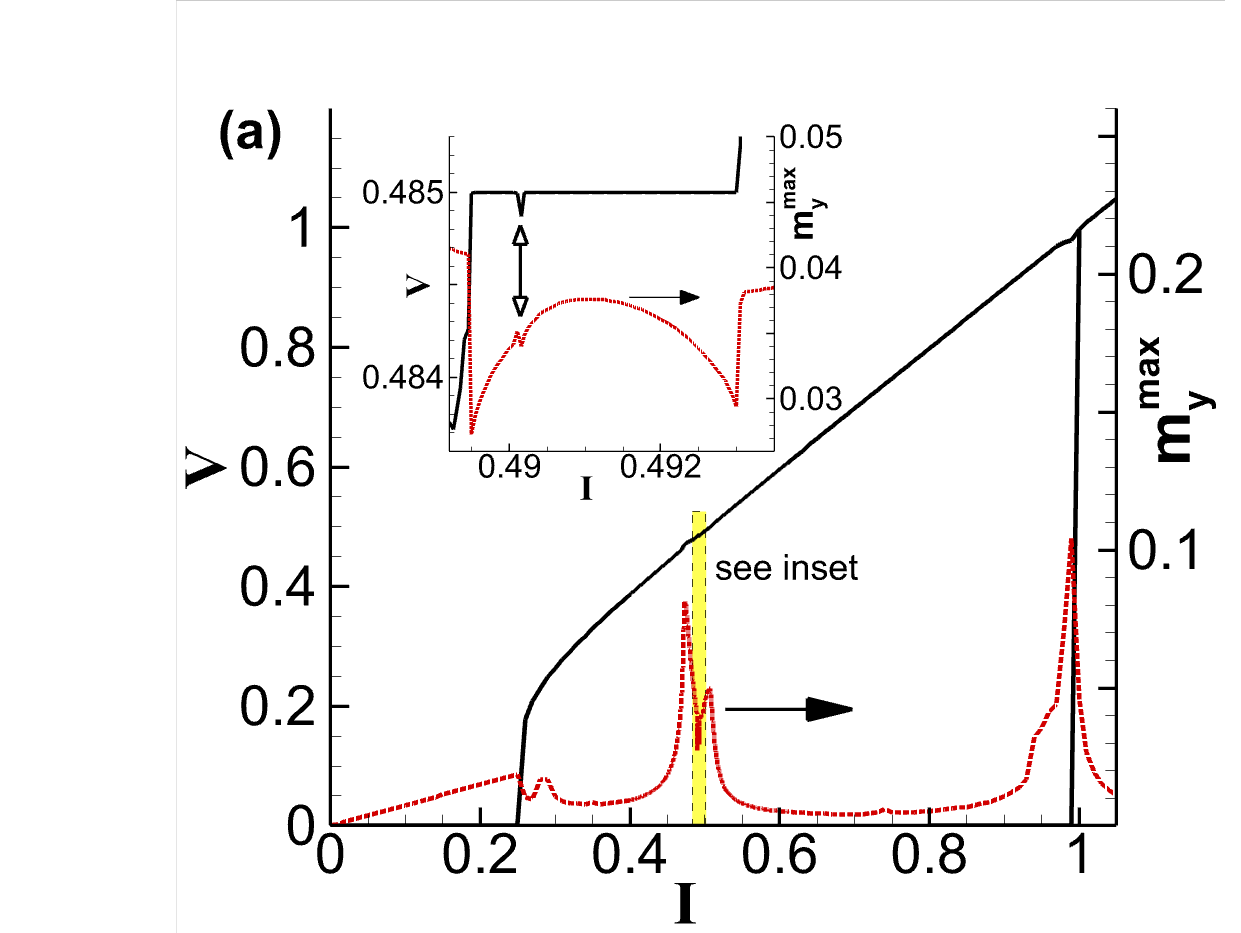}\\
	\includegraphics[width=0.8\linewidth]{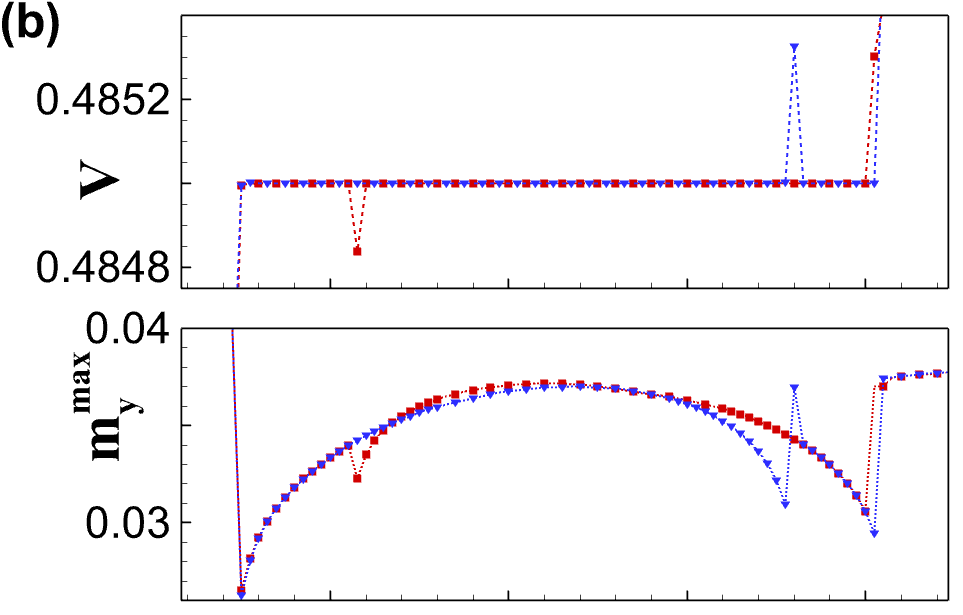}
	\includegraphics[width=0.8\linewidth]{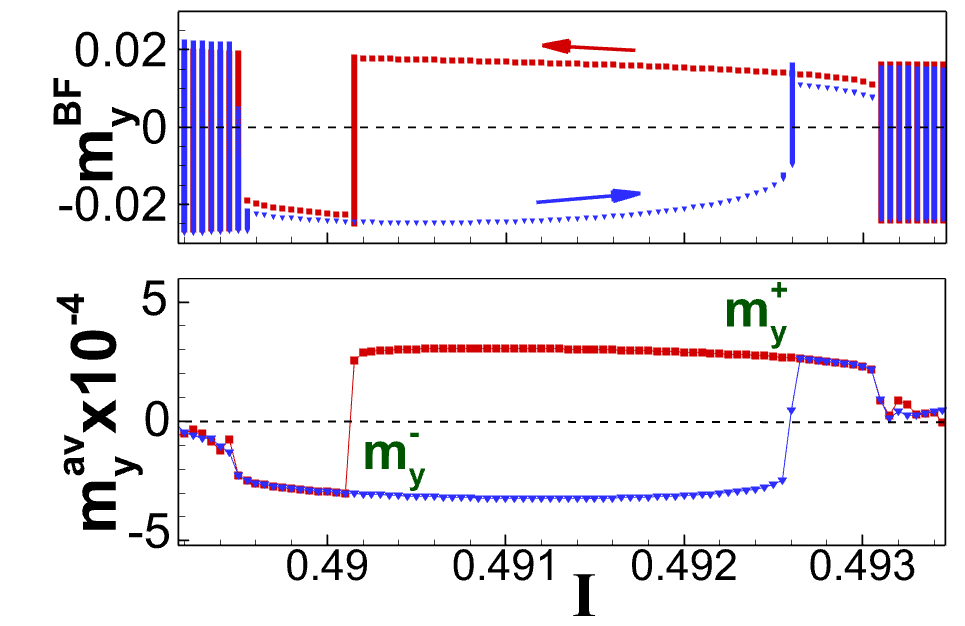}
	\caption{(a) Manifestation of chimera step in the current-voltage characteristic and $m_{y}^{max}(I)$ dependence for the $\varphi_{0}$ junction under external electromagnetic radiation with $A=0.005$ and $h_{R}= 1$. The current region where synchronization of Josephson oscillations and precession of the magnetic moment occurs is shown in the inset. (b) Enlarged part of $m_{y}^{max}(I)$, IV, magnetization bifurcation,  and $m_{y}^{av}(I)$ along the chimera step in the case of a decrease in current (red curve) and in the case of an increase of current (blue curve). All calculations are done at $r=0.4$, $\omega_R=0.485$.}
	\label{2}
\end{figure}

The origin for this surge is clarified in Fig.2(b) which shows together the $IV$- characteristic, $m_y^{max}(I)$, the bifurcation curves $m_y^{BF}(I)$ for $m_y(t)$ and the averaged value $m_{y}^{av}(I)$ in the vicinity of the chimera step for two directions in the bias current: decreasing (blue) and increasing (red).  The $m_y^{max}(I)$ dependence demonstrates two bubble structure: one bubble appears in the decreasing the bias current, another one in the increasing current. The bifurcation diagram calculations supports this conclusion. It is constructed by mapping for each current value a set of values $m_y$ taken through an integer number of radiation periods: $m_y(t_0 + 2n\pi / \omega_R)$, where n is an integer. This creates the Poincar\'e sections at each bias current's value. For the current value outside the step, these sets form vertical columns of finite height, since in this region the magnetization dynamics is not synchronized with the external field oscillations. In the current interval, corresponding to the chimera step (synchronization region), these sets show one point on the Poincar\'e section indicating one period motion (\textit{p1}-motion) \cite{MNashaat2022}, except for the current values corresponding to transitions between the bifurcation curves with positive and negative $m_y^{BF}$. We call the corresponding states as $m_y^+$ and $m_y^-$. Those states are manifested on $m_{y}^{av}(I)$ which shows also two curves with positive and negative $m_{y}^{av}$. For each bubble the average value could be positive or negative.  At transition  points, a restructuring of synchronization occurs, and we see Poincar\'e sections in the form of vertical segments. One transition point is in the direction of decreasing current, and the another is in the direction of increasing current. On the IV characteristic these points appear in the form of voltage spikes shown in the Fig. \ref{2}(b).

\begin{figure}[h!]
	\centering
	\includegraphics[width=\linewidth]{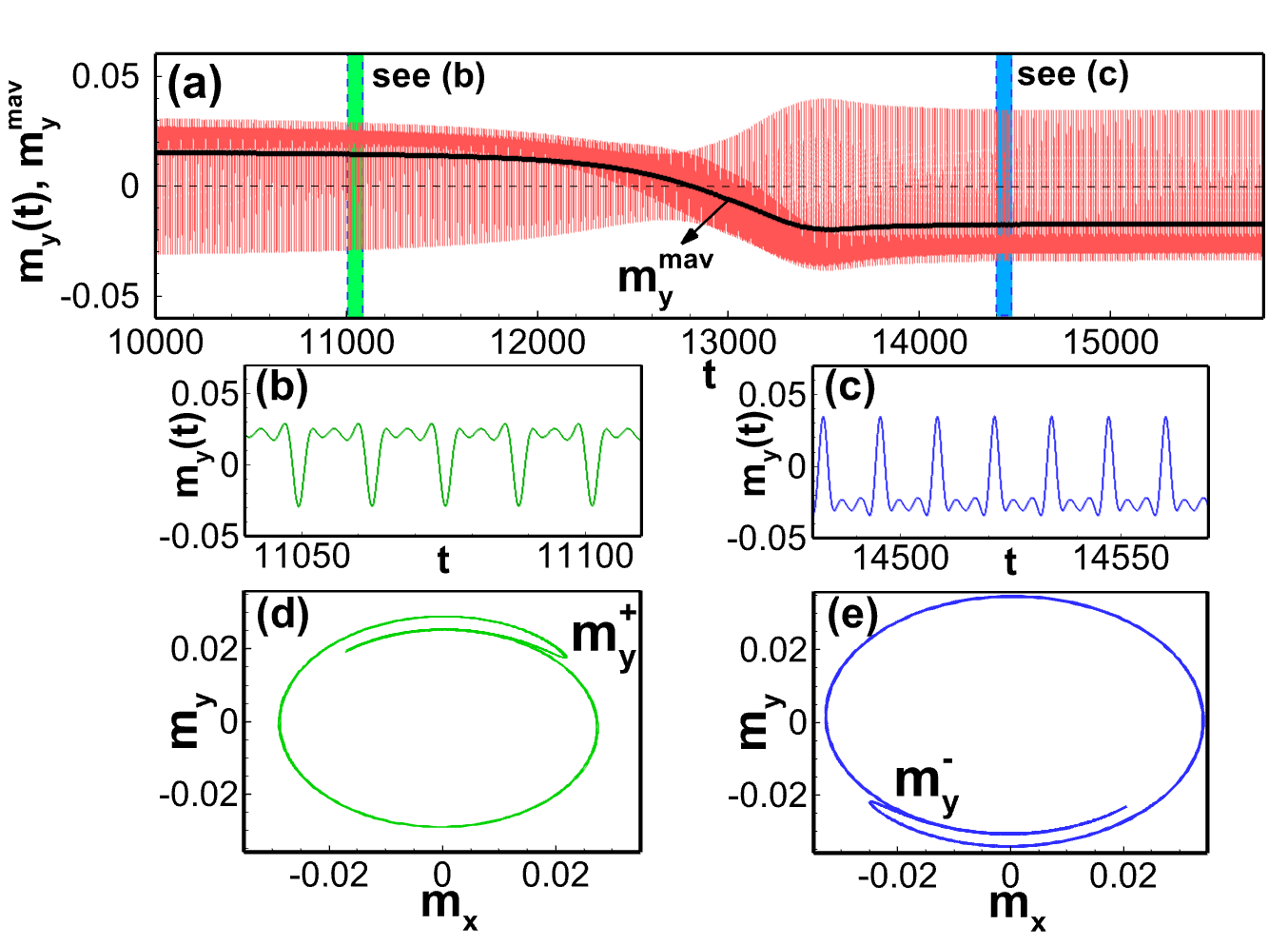}
	\caption{The change in the $m_{y}$ component dynamics in the case of the transition $m_y^+ \rightarrow m_y^-$ at $I= 0.49005$ (with decreasing current); (a) The corresponding time dependence in the transition region;  (b) and (c) represent the enlarged parts of $m_{y}$ time dependence before the transition (state $m_y^+$ ) and after (state $m_y^-$; (d) and (e) show the corresponding magnetization trajectories in the  plane $m_x - m_y$.}
	\label{3}
\end{figure}

To clarify the appearance of bubble structures in the $m_y^{max}(I)$ dependence and the occurrence of magnetization bifurcations, we studied the time dependence of $m_y$ at the bifurcation points. This dependence at a current value of $I=0.49005$, where magnetization bifurcation occurs as the current decreases, is presented in Fig.\ref{3}(a).

We observe a change in the nature of the $m_y$ dynamics during the transition process. Before the transition, the oscillations of $m_y(t)$ occur predominantly in the region of positive values of $m_y$, and after - mainly in the region of negative values, corresponding to the  dynamic states of magnetization $m_y^+$ and $m_y^-$, respectively. This fact is also demonstrated by the moving average value $m_y^{mav}$, shown by the black line, reflecting the states of $m_y^+$ before the bifurcation and $m_y^-$ after that. The averaging period when calculating $m_y^{mav}$ was chosen equal to the period of external radiation. Oscillations in the intervals highlighted in blue and green are shown in an enlarged scale in the Fig.\ref{3}(b) and (c), respectively. We note that at such transition, a phase difference shift of $\pi$ occurs in the time dependence of voltage and superconducting current.

The nature of the change in the dynamics of $m_y$ is also clarified by Fig. \ref{3}(d) and \ref{3}(e), which show magnetization trajectories in the $m_x - m_y$ plane in the states $m_y^+$ and $m_y ^-$, respectively. The characteristic curl of the motion trajectory (usually called among people the “mother-in-law’s tongue”) for the state $m_y^+$ is in the region of positive values of $m_y$, and in the region of negative values for the state $m_y^-$. It is important to note that these transitions appear on the IV characteristics in the form of spikes and, therefore, can be detected experimentally. It becomes possible to control the synchronization of magnetization in the $\varphi_{0}$ junction and change its state along the IV curve.

\begin{figure}[h!]
	\centering
	\includegraphics[width=0.8\linewidth]{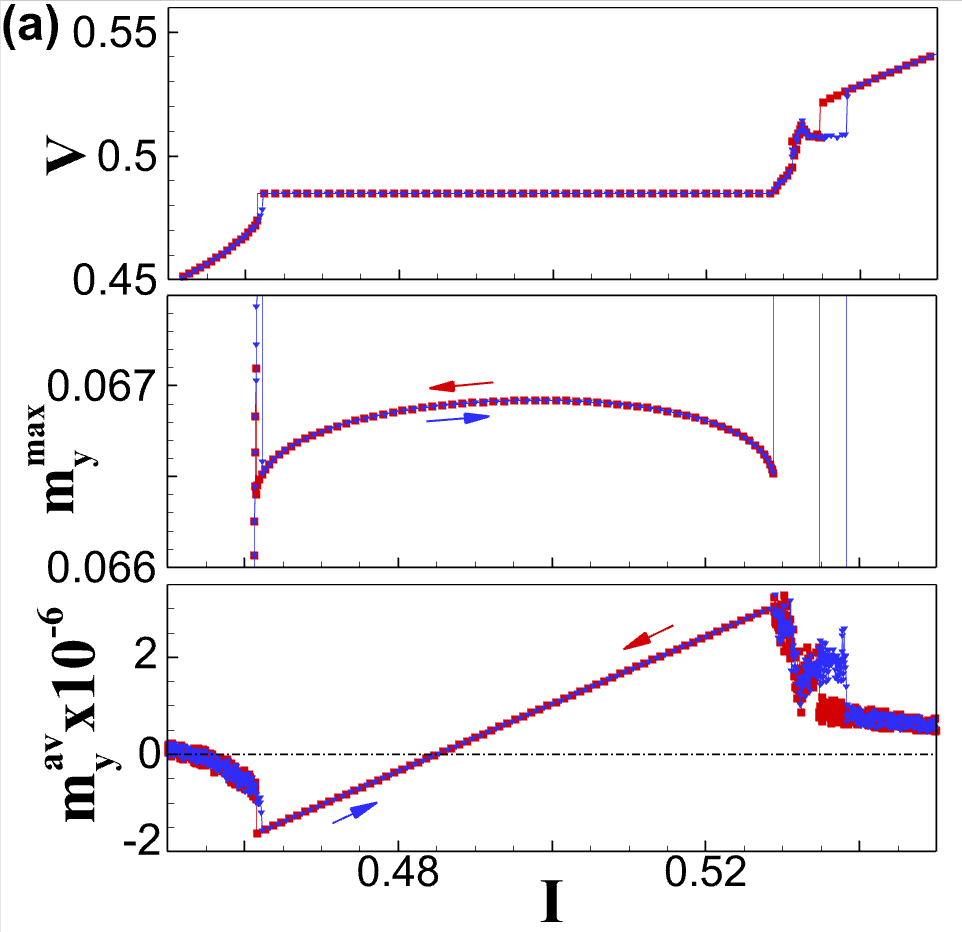}
	\includegraphics[width=0.8\linewidth]{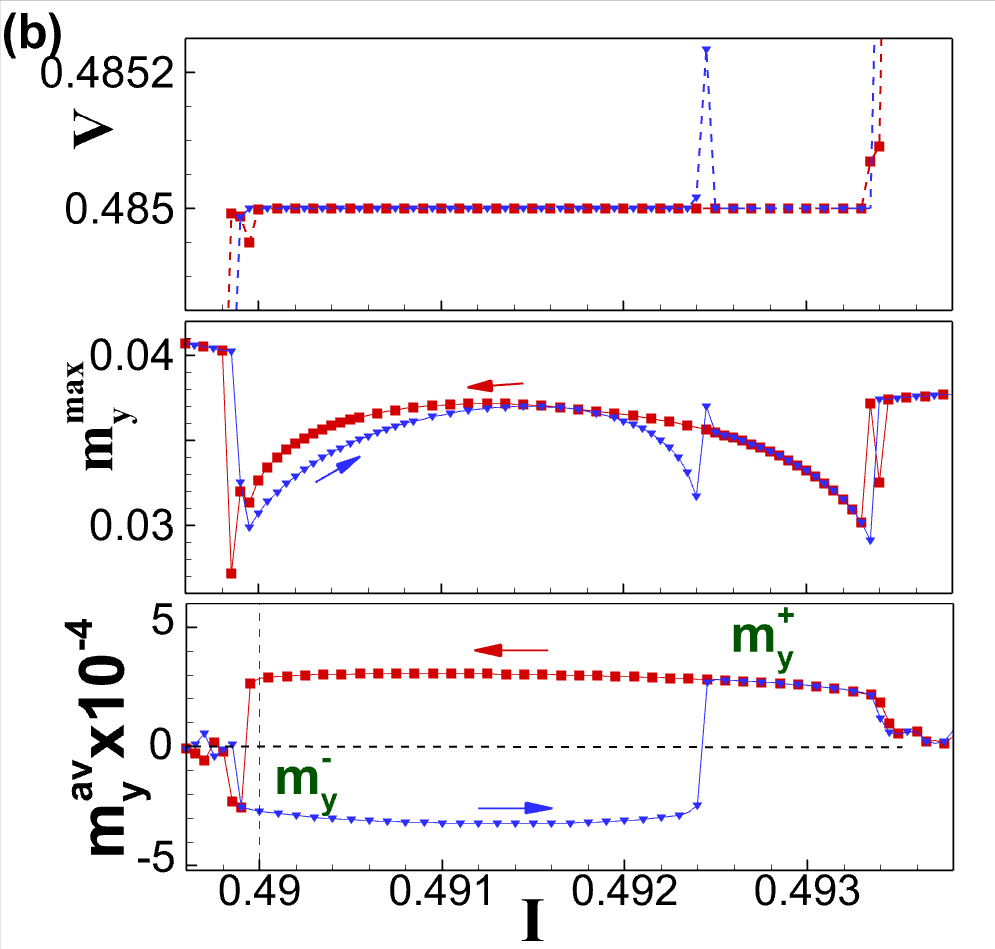}
	\caption{Comparison of two loop calculations  of $IV$-curve, $m_{y}^{max}(I)$ and average of $m_{y}^{av}(I)$ for:  (a) Shapiro step at $A=0.4$ and $h_{R}=0$, (b) Buzdin step at $A=0$ and $h_{R}=1$. All panels are done at $r=0.4$, $\omega=0.485$.}	
	\label{4}
\end{figure}

It would be interesting to compare the observed results for chimera step with the corresponded results for the Shapiro and Buzdin steps. In Fig.\ref{4}(a) we present an enlarged part of the $IV$-characteristics with Shapiro step ($h_{R}=0$).  As we see, for the Shapiro step $m_{y}^{max}(I)$ values has a very small variation, almost constant, along the step (see inset). The average value of $m_{y}(t)\approx10^{-6}$ along the Shapiro step is almost zero. In addition to this, the curves along the Shapiro step for the current simulation parameters are coincides for both direction of current (decreasing and increasing), i.e., two bubble stricture does not appear.

Results for the Buzdin step are presented in Fig.\ref{4}(b). Similar to chimera, the Buzdin step ($A=0$) demonstrates two bubble structure for $m_{y}^{max}(I)$ with the decreasing and increasing current. The intersection between the bubbles is manifested by surge on the Buzdin step in the $IV$-characteristics at $I=0.4899$ for decreasing current and at $I= 0.4925$ for increasing current (see supplemental materials \cite{supplemental}). Also,  $m_{y}^{av}(I)$ shows non-symmetrical hysteresis (see dashed horizontal line  in Fig.\ref{4}(b)). As we mentioned above, this feature open the way for experimental testing of the observed steps. We stress that in difference with chimera and Buzdin steps, the variation in $m^{max(min)}_{y}(t)$ values along the Shapiro step is much smaller than in the the cases of Buzdin and chimera steps, with average value $m_{y}^{av}(I)$ almost zero. Also, the dynamic charters and results of fast Fourier transformation (FFT) for $m_{y}(t)$ along the Shapiro step is different from the case of Buzdin step (see supplemental materials \cite{supplemental}).

\section{Switching between dynamical states}
\begin{figure}[h!]
	\centering
	\includegraphics[height=0.55\linewidth]{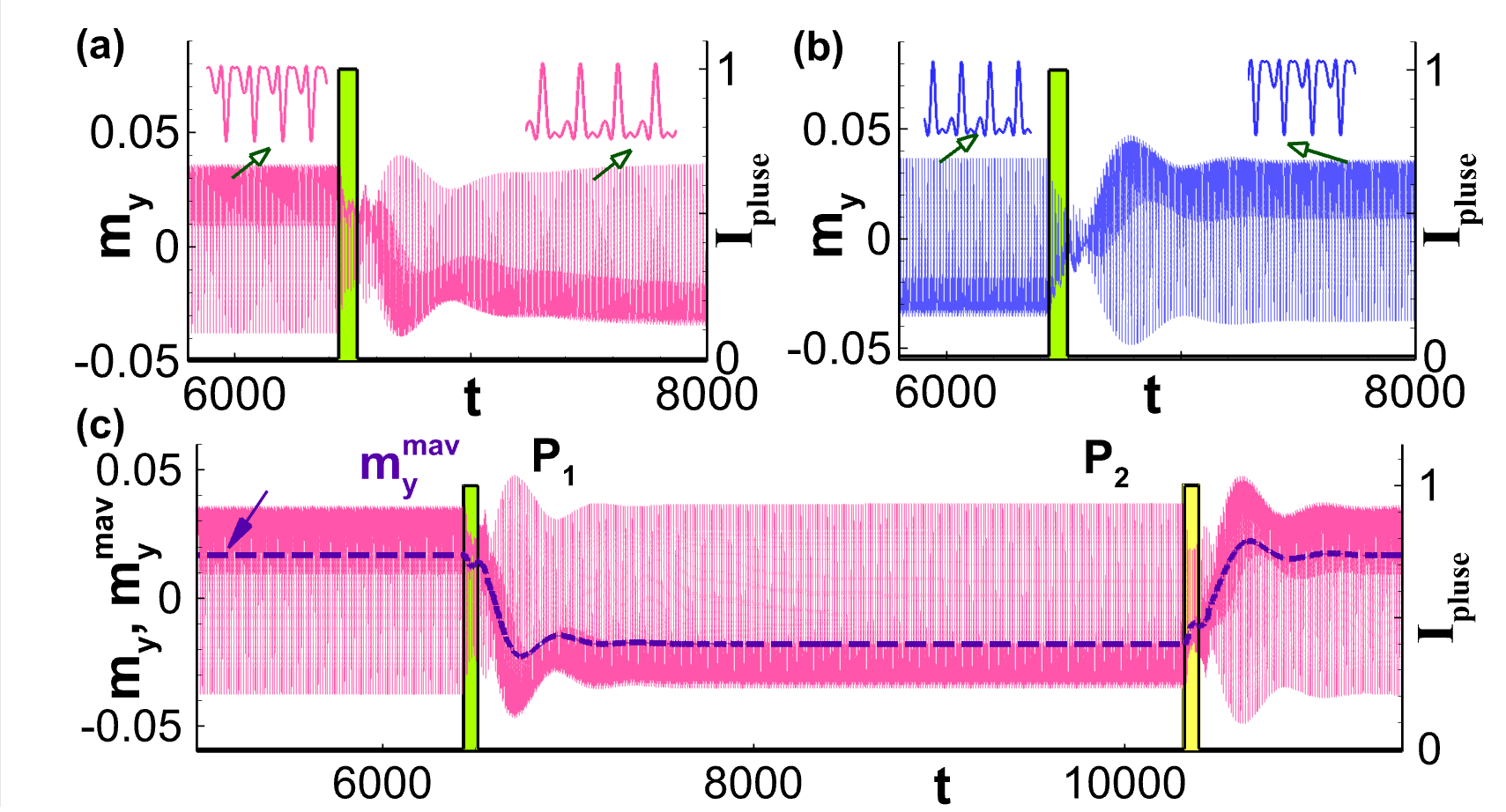}
	\caption{Magnetization dynamics for $m_{y}$ under rectangular pulse signal (a) decreasing current; (b) increasing current; (c) decreasing current with two successive pulses, the dashed line shows the moving average during these switching process. All panels are done with $r=0.4$, $h_{R}=1$ and $A=0.005$. }
	\label{5}
\end{figure}

Existence of two different magnetization states $m^{+}_y$ and $m^{-}_y$ in bias current interval, corresponded to the chimera or Buzdin step, creates a series of novel interesting phenomena and applications in superconducting spintronics. In particular, one of them is an implementation of controllable switching between these two states. Because the states are determined  by bias current direction, one obvious way is the switching between them by sweeping or changing the current direction. Another way is to apply a current pulse similar to what was already discussed for magnetization reversal in the $\varphi_0$ junction \cite{shukrinov-apl17,Bobkova2020}. Here we demonstrate such possibility of switching  by bias current pulse of rectangular form in Fig.\ref{5}.

We have chosen an arbitrary value of bias current $I=0.492$, at which the both two states of $m_{y}$ are realized for increasing and decreasing bias current process (see Fig.\ref{2}). Switchings are demonstrated in Fig.\ref{5} (a) and (b) for decreasing (transition $m_{y}^{+} \rightarrow m_{y}^{-}$ ) and increasing current direction ($m_{y}^{-} \rightarrow m_{y}^{+}$), respectively.  With the given simulation parameters, the switching between the two states occurs for pulse width $6T$ and pulse amplitude $I_{pulse}=1$, where $T$ is the period of the external electromagnetic field, and the pulse amplitude is normalized to $I_{c}$. The switching with a changed pulse parameters are shown in sections "B" of the supplemental material \cite{supplemental}, demonstrated it for pulse width $8T$ and pulse amplitude $I_{pulse}=0.8$. So, the transition between states can be controlled by the pulse and model parameters, by analogy of magnetization reversal discussed in Ref.\cite{Bobkova2020}, Moreover, we can also control the switching between the two states in fixing the current direction and just apply successive pulses as demonstrated in Fig.\ref{5}(c), which shows the switching from $m_{y}^{+}$ state to $m_{y}^{-}$ one by applying the first pulse ($P_{1}$), then after some time the second pulse ($P_{2}$) switch the state  $m_{y}^{-}$ to $m_{y}^{+}$. The line shows a moving average during these switching process.

\section{Variation of hysteresis by spin-orbit coupling and radiation parameters}

Another interesting and important phenomena related to the creation of the two different magnetization states $m^{+}_y$ and $m^{-}_y$ directly follows from the results, presented in Fig.\ref{2}. Sweeping current along the chimera step, decreasing it in the first loop of sweeping and increasing in the second one, demonstrates a hysteretic behavior in the dependence $m_y^{max}(I)$. It's also clearly pronounced in the simulations of the bifurcation $m_y^{BF}(I)$  and averaged $m_y^{av}(I)$ dependencies. At chosen junction and electromagnetic field parameters indicated in the caption to Fig.\ref{2}, transition between states $m_y^+ \rightarrow m_y^-$ (red color) occurs at $I=0.4901$, while  $m_y^- \rightarrow m_y^+$ (blue color)  happens at  $I=0.4925$, forming a hysteresis loop with width equal to $\Delta I=0.0024$.

\begin{figure}[h!]
	\centering
	\includegraphics[width=\linewidth]{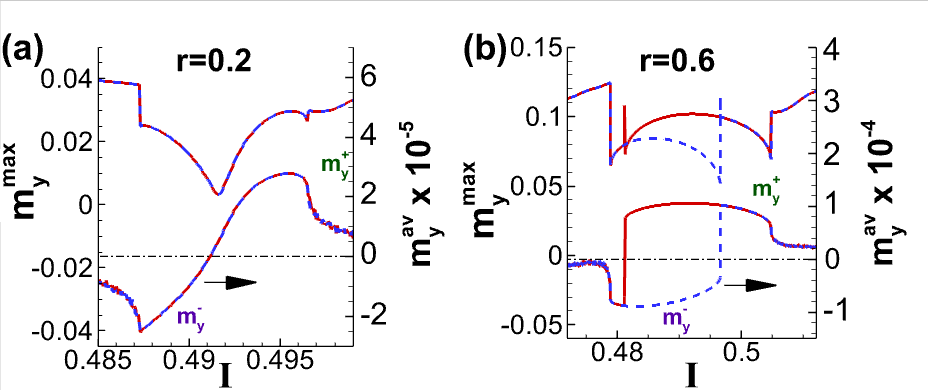}
	\caption{Variation of the hysteresis region with a change in the spin-orbit coupling parameter. (a) Results of two loop calculations of $m_y^{max}(I)$ and $m_y^{av}(I)$  at $r=0.2$; (b) The same at $r=0.6$. }
	\label{6}
\end{figure}

Figure \ref{6} demonstrates the effect of spin-orbit interaction (SOI) on the hysteresis. It demonstrates $m^{max}_{y}$ and $m^{av}_{y}$ as functions of bias current at two values of SOI parameters.

As we see, at a small value of SOI parameter ($r=0.2$), the $m_y^{max}(I)$ curves coincide for both directions of current changing, i.e. the hysteresis is absent. The dependence $m^{av}_{y}(I)$ manifests the both two states of $m_{y}^{(\pm)}$ which correspond to the right and left halves of bubble in $m_y^{max}(I)$. As we noted in Ref. \cite{shukrinov-prb24},   the width of Buzdin and chimera steps are growing crucially with increase in spin-orbit coupling. Compare the results for hysteresis width $\Delta I$ at different SOI parameter, we find that it grows crucially with an increase in $r$ as well. In particular, at $r=0.4$ the hysteresis width $\Delta I=0.0024$, while at $r=0.6$ it is $\Delta I=0.0156$. But relatively to the step width, the hysteresis width does not change essentially.

\begin{figure}[h!]
	\centering
	\includegraphics[width=0.8\linewidth]{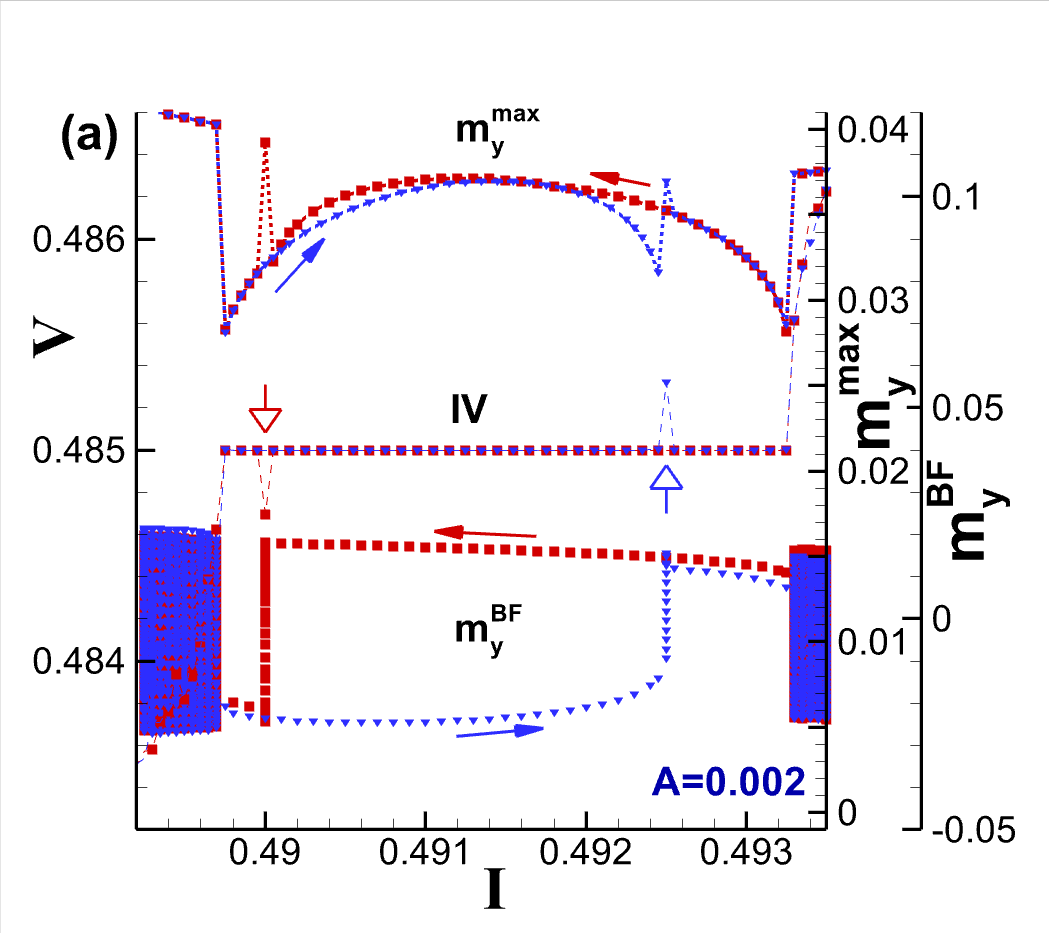}\\\includegraphics[width=0.8\linewidth]{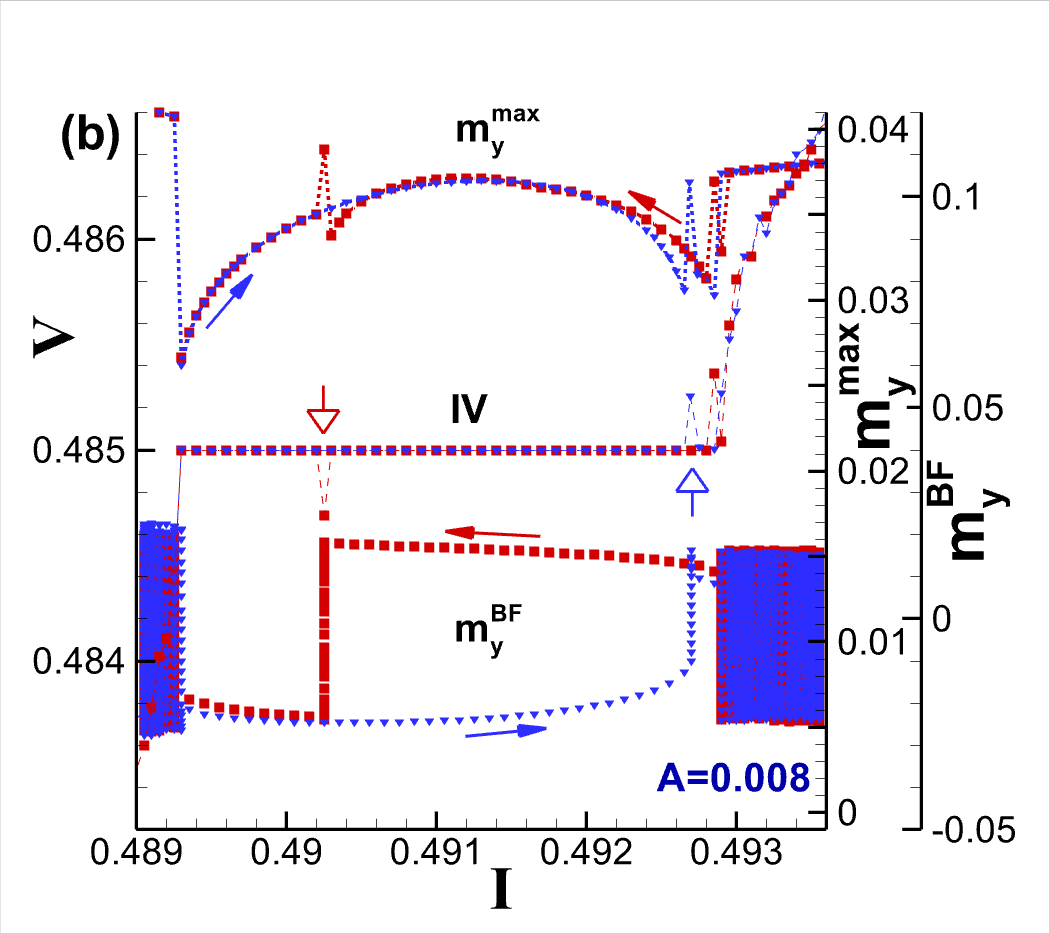}
	\caption{Effect of external radiation on the hysteresis features. (a) Results of two loop calculations of $IV$-characteristics, $m_y^{max}(I)$ and $m_y^{BF}(I)$  at  $A=0.002$;  b) The same at $A=0.008$.}
	\label{7}
\end{figure}

Figure \ref{7} shows the effect of slight variation of radiation electric component on the hysteresis features.
Taking into account  also results presented in Fig.\ref{2}, we can make a conclusion, that an increase in $A$ in the interval (0.002-0.008) leads to shifting of the hysteresis region from the right part of step to the left one of the chimera step. Transition $m_y^-  \rightarrow  m_y^+$ with an increasing in the bias current is observed at the edge of the step. The width of the hysteresis region does not change practically. We stress ones more that the spikes in the presented $IV$-characteristics correspond to the transition between the two states of $m_{y}$ and indicate the size of hysteresis region. The $IV$-characteristic of Josephson junction can be measured experimentally, so, the observing spikes in it, we indicate a direct way to investigate the transition points between two bubbles corresponded to different states of magnetization and hysteric features of the system experimentally.

\section{Results for one-signal model}
Results of calculations in one-signal model are less expressive, but they are qualitatively consistent with the described observations. To stress this fact and prove the correctness of our conclusions, we present here two examples, demonstrating the bifurcation along the chimera step and two dynamical states of $m_{y}^{+(-)}$  in one-signal model. As in Ref.\cite{shukrinov-prb24} we use the same set of model parameters: the critical current density $J_{c}= 10^{5}$ (A/$cm^{2}$), the Josephson junction area ($0.1 \times 0.1$) $\mu m^{2}$, the normal resistance R=1 $\Omega$, the characteristic frequency $\omega_c=30$ GHz, and relative magnetic permeability $\sim 10 ^{5}$. Using the obtained relation between $h_R$ and $A$ :
\begin{eqnarray}
\label{formul}
h_{R}= \frac{\gamma I_{c} A}{\nu^{3/2} \omega_{c}} \sqrt{\frac{2 R}{S \epsilon},}
\end{eqnarray}
where $\nu=1/\sqrt{\mu\epsilon}$, and $ S $  is the area of the JJ, considering the magnetic permeability within $10^{2} \sim 10 ^{5}$, we might have the values $A=0.5$ and $h_R$ in the interval (0.002 - 0.149) for power $P = 10^{-10}$ Watt.  The power $P = 10^{-8}$ Watt leads to the values $A=10$ and $h_R$ in the interval (0.017 - 2.982).

\begin{figure}
	\centering
	\includegraphics[width=0.9\linewidth]{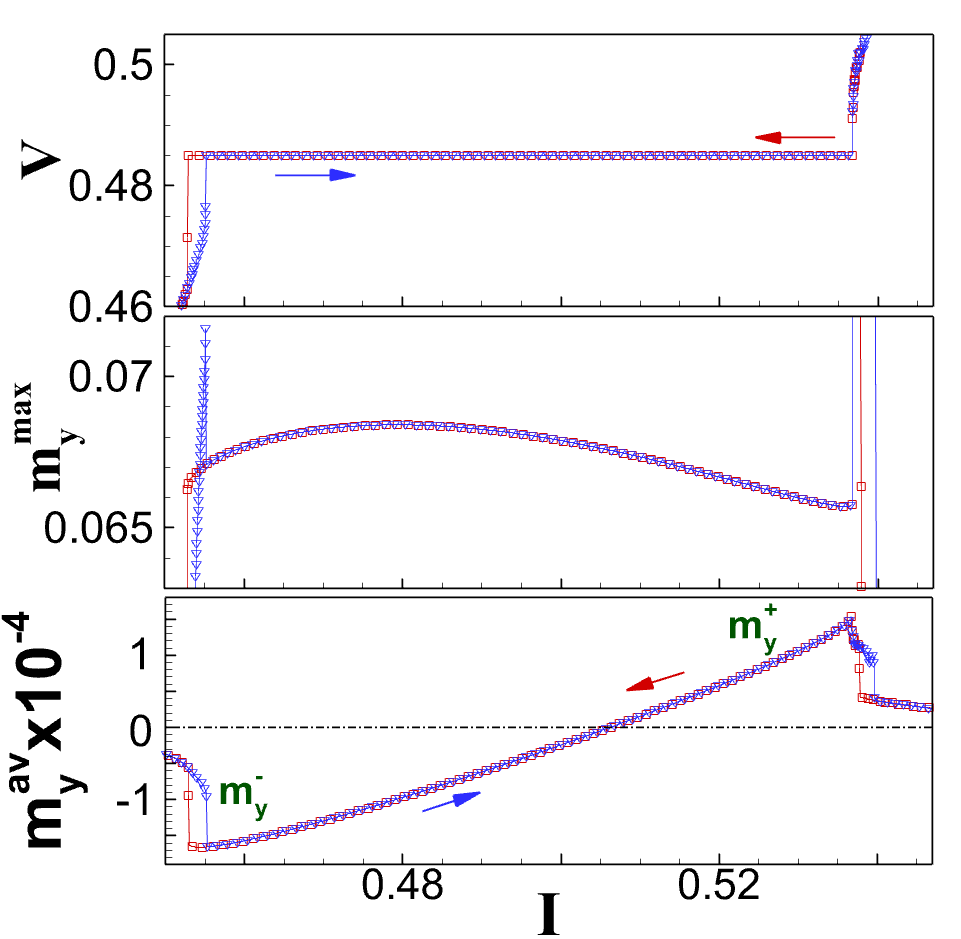}
	\caption{An enlarged part of $IV$-characteristic, $m_{y}^{max}(I)$, and $m_{y}^{av}(I)$ along the chimera step at $A=0.5$, $h_{R}=0.149$, $r=0.4$, and $\omega_{R}=0.485$.}
	\label{8}
\end{figure}

As a first example, we consider the case with $A=0.5$ and the corresponding $h_{R}=0.149$. Figure \ref{8} shows an enlarged part of the $IV$-characteristic with chimera step, and the corresponding dependencies  $m_{y}^{max}(I)$ and $m_{y}^{av}(I)$ for both current directions. As we see, no hysteresis is demonstrated at those values, while we have continuous transition between the two dynamical states of $m_{y}^{+(-)}$ by changing the current along the step.

In the second example, assuming critical current density $J_{c}= 10^{5}$ (A/$cm^{2}$), Josephson junction area ($0.126 \times 0.126$) $\mu m^{2}$, and resistance R=1 $\Omega$, we get the characteristic frequency $\omega_c=48$ GHz. With microwave power $P\approx  3.8\times 10^{-7}$ Watt, and for relative permeability $10^{4}$, according to the relation (\ref{formul}), we find $h_{R}=1.64$  for $A=38.9$ .

\begin{figure}
	\centering
\includegraphics[width=0.9\linewidth]{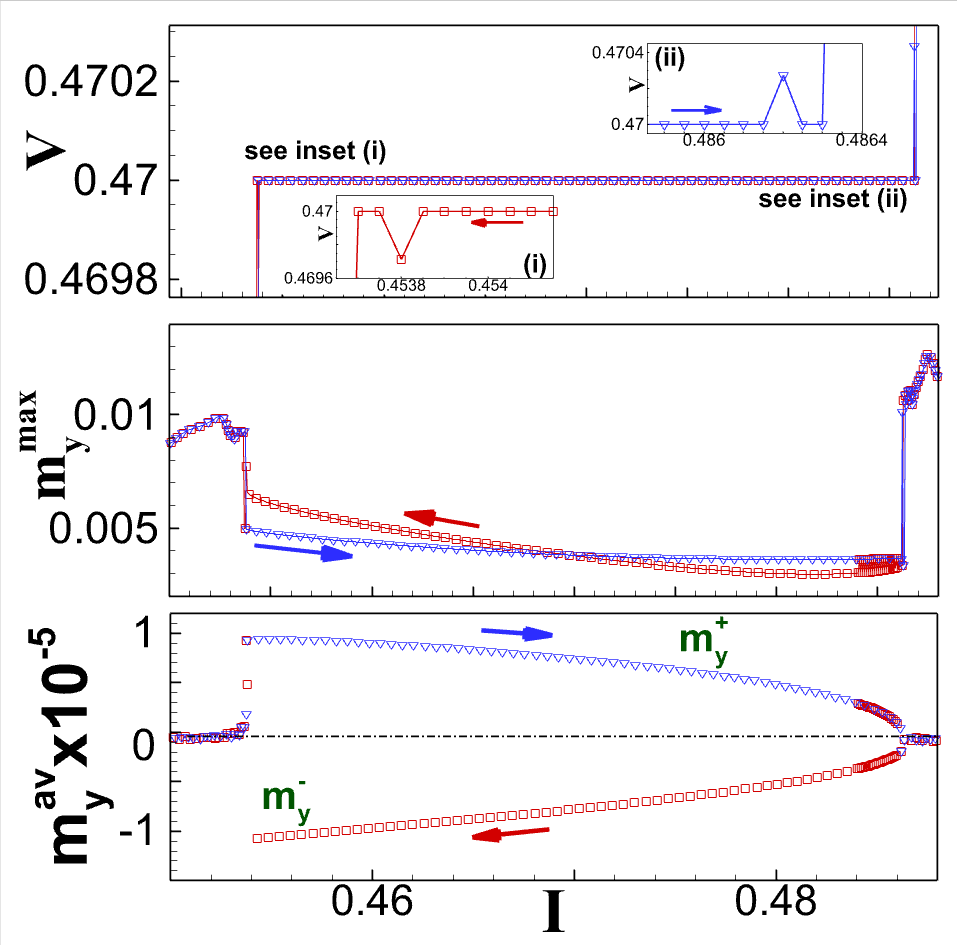}
\includegraphics[width=0.9\linewidth]{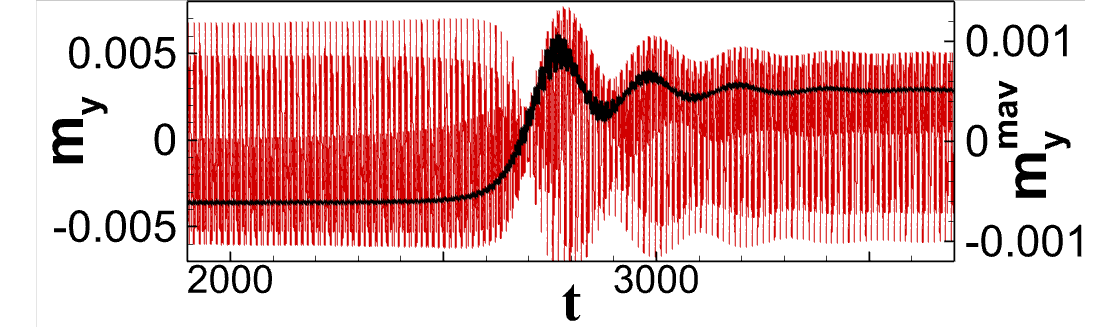}
	\caption{An enlarged part of $IV$-characteristic, $m_{y}^{max}(I)$, and $m_{y}^{av}(I)$ along the chimera step at $A=38.9$, $h_{R}=1.64$, $r=0.2$, and $\omega_{R}=0.47$.}
	\label{9}
\end{figure}

The results of calculations with these parameters are illustrated in Fig.\ref{9}.

Here, $m_{y}^{av}(I)$ demonstrates hysteresis along the chimera step, and we see the transition between the two states $m_{y}^{\pm}$ at $I=0.45380$ for decreasing the current direction. In addition to this, the moving average demonstrates the existence of the two states of $m_{y}$. Also, it can be seen that on the right side of the chimera step, where the difference between the $m^{\pm}$ states is small, an instability occurs and the average value $m_y{av}$ fluctuates between these two states. The fluctuations disappear when the difference between the two branches become large like in the left side of the step. This feature and its dependence on the model parameters will be investigated in detail somewhere else and it is out of the scope of the current work.

On the other hand, as was shown in Ref.\cite{Efetov2011},  it is possible to apply microwave field along with ac magnetic one. In this case, we would have two-signal model and $A$ is decoupled with $h_{R}$. Thus, it is possible to tune the value of $h_{R}$ to observe the transition between the two states of $m_{y}$. Also, this  transition is manifested in phase shift of $\pi$ for the supercurrent and voltage for chimera step (or Buzdin step if we apply only magnetic field).

\section{Phase shift on the chimera step}
\begin{figure}[h!]
	\centering
	\includegraphics[width=0.9\linewidth]{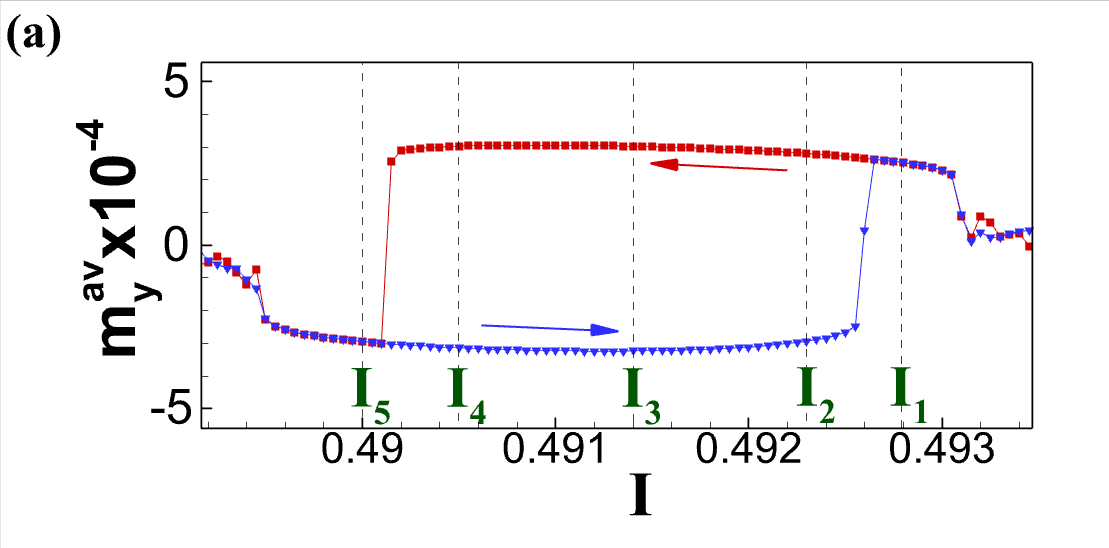}
	\includegraphics[width=\linewidth]{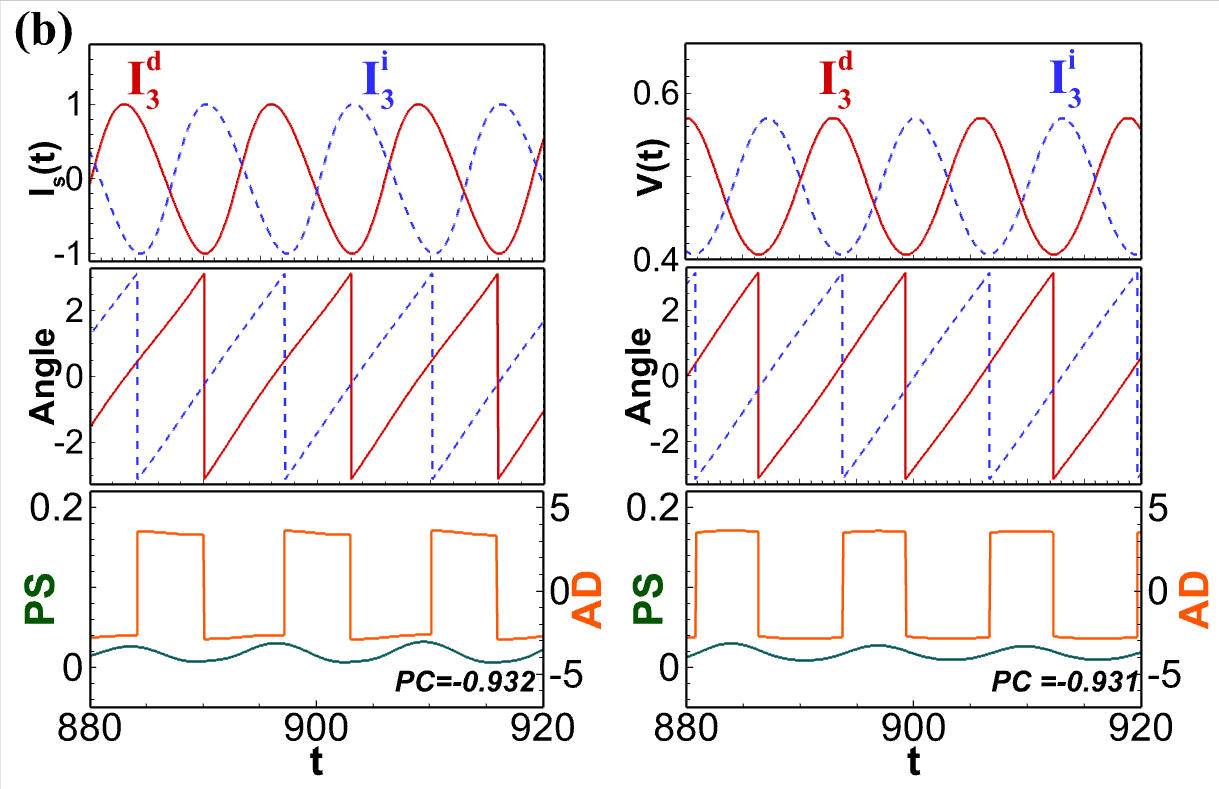}
	\caption{Demonstration of the phase shift in the chimera step. (a) An enlarged part of the $m_{y}^{av}(I)$ in chimera step for two loop calculations. The dashed vertical lines indicate the positions at which we have investigated the temporal dependence of the supercurrent and voltage for decreasing (d) and increasing (i) current directions. (b) Superonducting current and voltage signals analysis for the pair ($I^{d}_{3}$:$I^{i}_{3}$). The temporal dependence of angle, phase synchrony and angle difference is demonstrated. The simulation parameters are same as for Fig.\ref{2}(b).}
	\label{10}
\end{figure}

An additional interesting property of the investigated system follows from the  analysis of the temporal dependence of superconducting current and voltage at different points along the chimera step shown in Fig.\ref{2}(b).  We choose several points fixed by dashed lines in Fig.\ref{10}(a) that lie on different branches of $m^{av}_{y}$ and designated as $I_n$.
	
In Fig.\ref{10}(b) we present the results of signal analysis for supercurrent at points $I^{d}_{n}$, $n=1,2,3,4$ for decreasing (d) and increasing (i) current direction. To analyze the temporal dependence of the suppercurrent, we perform Hilbert transformation \cite{Poularikas2018} of the suppercurrent signal to find the phase. In this case, we convert the time signal into an analytic signal, such that the signal become a vector of complex numbers whose magnitude is constant and whose phase is changing in synchronization with the original signal. This allows us to find the instantaneous phase lag between two signals. Also, we calculate the instantaneous phase synchrony (PS) which measures the phase similarities between signals at each time. PS can be calculated by subtracting the angular difference from 1, i.e., $PS=1-\sin((\mid\theta_{i}-\theta_{j}\mid)/2)$ where $i$ and $j$ represent different points in Fig.\ref{10}. If the two signals line up in phase their angular difference becomes zero.

In addition to this, we calculate also, the angle difference ($\delta \theta$) between two suppercurrent signals at different position, and Pearson correlation (PC) which measures how two continuous signals co-vary over time and indicate the linear relationship as a number (linear correlation). It has a value between -1 to 1, with a value of -1 meaning a total negative linear correlation, 0 being no correlation, and + 1 meaning a total positive correlation. If the phase shift between two signal is $0$, or any multiple of 2$\pi$, the waves are identical and so the correlation should be 1. When the phase shift is $\pi$, or an odd multiple of $\pi$, the two waves are perfectly out of phase, and so the correlation should be -1. In between these extremes the correlation oscillates.

We illustrate an example for the temporal dependence of the suppercurrent (voltage), instantaneous angle in radiant, PS and PC for the pair ($I^{d}_{2}$:$I^{i}_{2}$). As shown in Fig.\ref{10}(b), if the two points lie on two different bubble, a phase shift $\approx \pi$ appears between them. This is confirmed by negative value of PC. In Table.\ref{table} we calculate the PS and PC for all points shown in Fig.\ref{10}(b). As it is illustrated, if the two points lie on two different bubble, a phase shift close to $\pi$ with negative PC value. However, the phase shift and Pearsons correlation takes intermediate values between the points on the same current direction (see section "C" of the supplemental material \cite{supplemental} ), this case is similar to trivial SIS Josephson junction.

\begin{table}[!htbp]
	\centering
	\caption{\label{table} Measurements of the Phase shift (PS) in degree and Pearsons correlation (PC) for several points shown in Fig.\ref{2}(b). Points $I^{d(i)}_{1}$ corresponds to $I=0.4928$, $I^{d(i)}_{2}$ corresponds to $I=0.4923$, $I^{d(i)}_{3}$ corresponds to $I=0.4915$, $I^{d(i)}_{4}$ corresponds to $I=0.4905$ and $I^{d(i)}_{5}$ corresponds to $I=0.49$. We use (d) and (i) for decreasing and increasing current direction.}
	\begin{booktabs}{
			colspec = {crrrr},
			cell{1}{2,4} = {c=2}{c}, 
		}
		\toprule
		Pair &   $I_{s} (t)$ &      &  V(t) &      \\
		\midrule
		&    PS &  PC &  PS & PC \\
		\cmidrule[lr]{2-3}\cmidrule[lr]{4-5}
		$I^{d}_{1}$:$I^{i}_{1}$  &  0$^{o}$ & 1 &  0$^{o}$ & 1 \\
		$I^{i(d)}_{1}$:$I^{i}_{2}$  &  176$^{o}$ &  -0.888 & 179$^{o}$& -0.898 \\
		$I^{d}_{3}$:$I^{i}_{3}$  &  177$^{o}$  & -0.932 &  181$^{o}$ & -0.931  \\
		$I^{d}_{4}$:$I^{i(d)}_{5}$  &  175$^{o}$  & -0.872  &   179$^{o}$ &  -0.876  \\
		$I^{d}_{5}$:$I^{i}_{5}$  &   0$^{o}$ & 1 & 0$^{o}$  & 1 \\
		\bottomrule
	\end{booktabs}
\end{table}

To understand the origin of the phase shifts which are demonstrated in Table.\ref{table}, let us consider the difference between the suppercurrent at two points on the step is given by $\delta I_{s}=\sin(\phi_{1}-rm_{y_{1}}) - \sin(\phi_{2}-rm_{y_{2}})$. In general, $\phi_{1}\neq \phi_{2}$ on the step, because the phase may take any value, however $d\varphi/dt=$ must be constant. Now, if $h_{R}=0$, $G=0$ and $r=0$, the IV curve will demonstrate only Shapiro step (as in single JJ). In this case, the suppercurrent will not show any phase shift by switching current direction at fixed current value. However, a phase shift appears if we sweep the current along the Shapiro step. In chimera step the situation is very interesting. A phase shift of $180^{o}$ appears by switching the current direction due to the appearance of the two states of $m_{y}^{\pm}$ (i.e., if $m_{y_{1(2)}}$ changes sign). Then the suppercurrent for the two points become out of phase, which is the case for two points on different bubble.

So, by switching the current direction (or apply current pulse), our results can provide an experimental way to observe the two states of magnetic moment and measure its hysteresis area by recording the suppercurrent (voltage) temporal dependence on each current step and then calculate the phase shift for both direction of current. It is known that, a phase shift oscillator produces an output which is $180^{o}$ out of phase to the input \cite{Bishop2011}. The above results may open the way to fabricate spintronic device and operate it as a phase shift oscillator in the region of chimera step.

\section{Conclusions}
The interaction of the magnetic field of microwave radiation with the magnetic moment of the $\varphi_0$ Josephson junction leads to a number of unique resonant and synchronization phenomena. Particularly, due to the coupling of superconductivity and magnetism in this system, two mechanisms of synchronization of Josephson oscillations and magnetic precession are appeared. The precession of the magnetic moment of the ferromagnetic layer, caused by the magnetic component of the external radiation, leads to the creation of the Buzdin and chimera steps in the $IV$-characteristics of the $\varphi_0$ junction \cite{shukrinov-prb24}.

As a summary, we have explained the origin of bubble structure in the bias current dependence of magnetization in SFS $\varphi_0$ Josephson junction under external electromagnetic radiation. The appearance of two different dynamic states in the synchronization region chimera step of magnetic precession and Josephson oscillations on the $IV$-characteristics were demonstrated. These states are differed by temporal dependence of magnetic moment, and there are the phase shift of $\pi$ in the suppercurrent and voltage dynamics. Moreover, the transition between these states leads to novel type of hysteresis which is strongly depend on the spin-orbit coupling and radiation parameters. We have demonstrated that the transition and hysteresis between theses locking states are reflected on the bifurcation diagram and $IV$-characteristics. So, the obtained results might be tested experimentally. We consider that the presented results will be useful for quantum computing devices, and, in particular, for the phase shift oscillator.

\section{Acknowledgments}
The authors are grateful to A. Buzdin, Y. Fominov, J. Tekic, I. Rahmonov, K. Kulikov, T. Belgibaev, G. Ovsyannikov, K. Constantinian for fruitful discussion of some results of this paper. The work was carried out with the financial support of Project No. 24-21-00340 of Russian Science Foundation (RSF), numerical modeling was carried out with the financial support of Project  No. 22-71-10022  of RSF. Yu. M. Shukrinov and  M. Nashaat acknowledge the financial support from  collaborative project ASRT, Egypt – JINR, Russia. Special thanks to the Alexandria Library (Egypt), BLTP and the heterogeneous computing platform HybriLIT, LIT, JINR, Russia for the HPC servers.

\bibliographystyle{unsrtnat}

\newpage
\thispagestyle{empty}
\mbox{}
\newpage

\break

\newpage
\onecolumngrid

\section*{Supplemental material for "Buzdin,  Shapiro and Chimera Steps in  $\varphi_0$ Josephson Junctions. II. Bifurcation, Switching, and Hysteresis."}
\setcounter{figure}{0}
\setcounter{table}{0}

\subsection{Bifurcation along the Shapiro and Buzdin steps}
\begin{figure}[h!]
	\centering
	\includegraphics[width=0.45\linewidth]{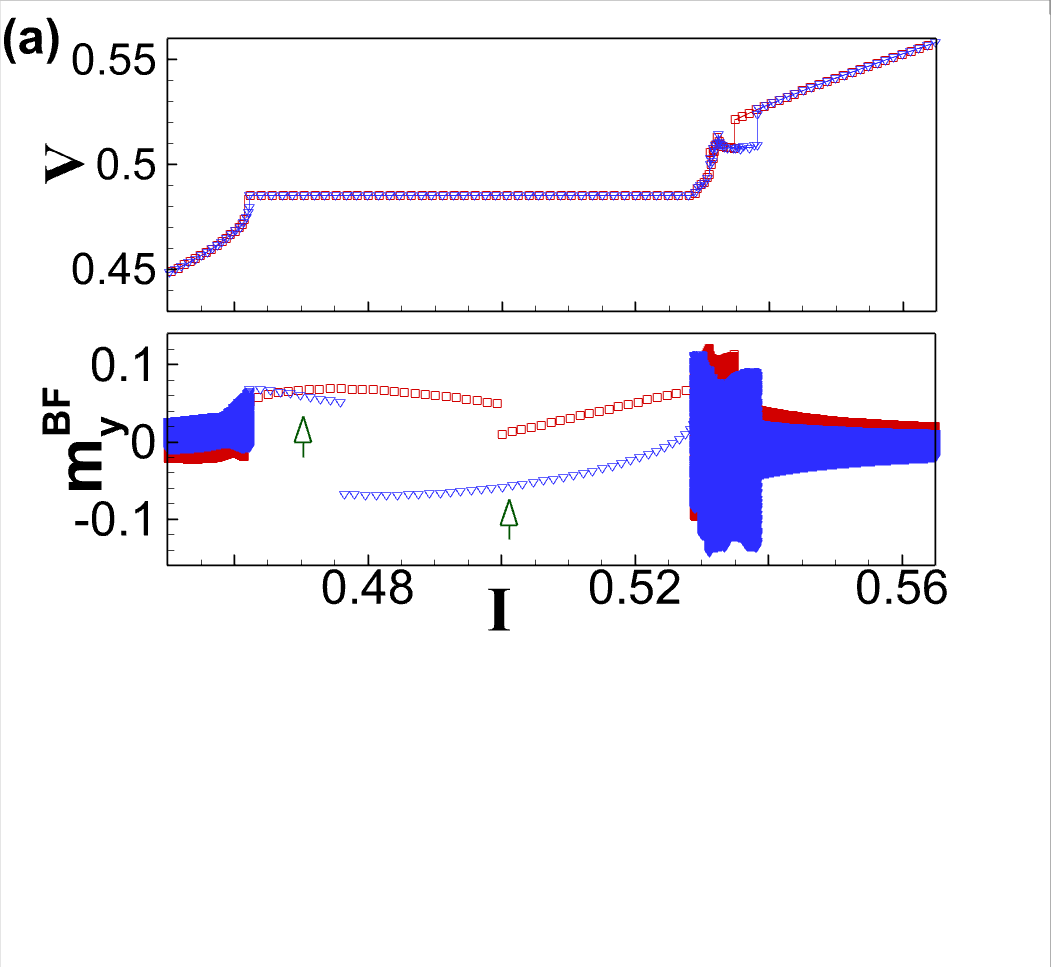}
	\includegraphics[width=0.45\linewidth]{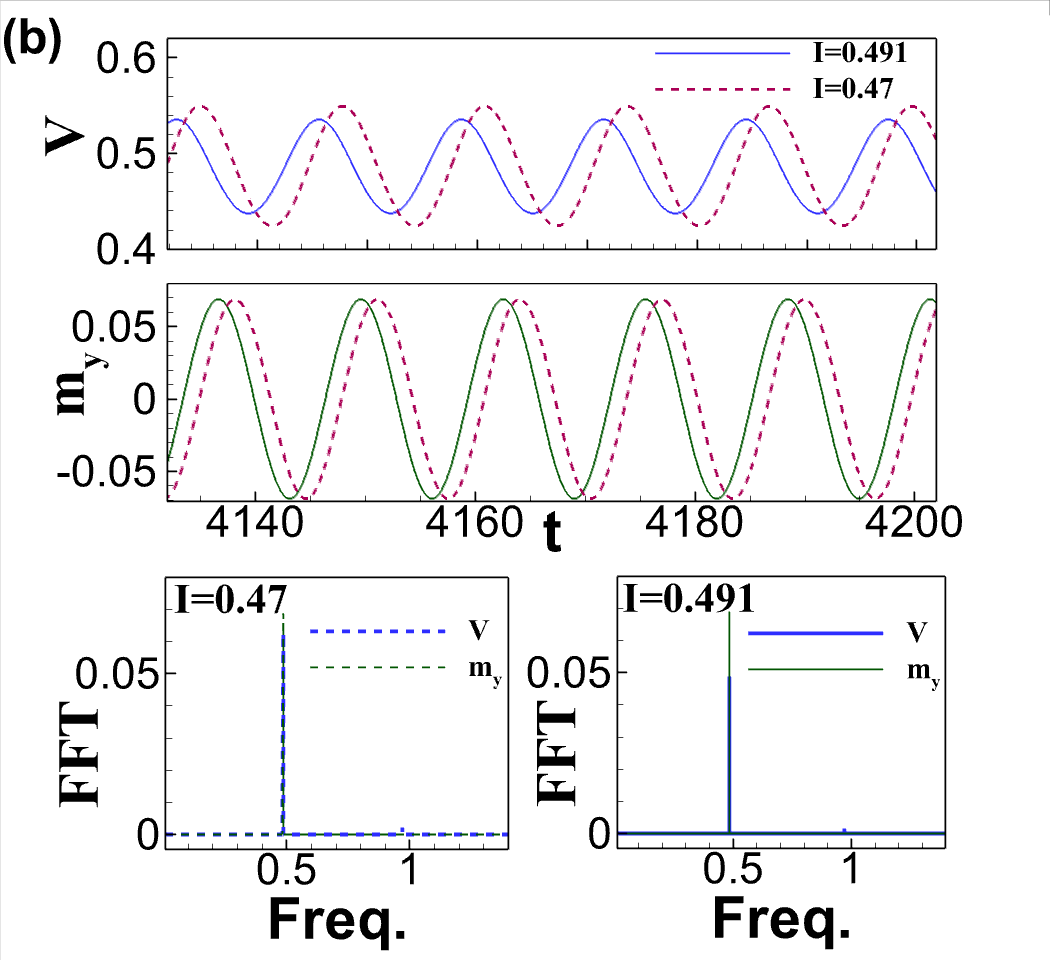}
	\caption{ (a) Enlarged part for Shapiro step with two loop calculation for $IV$-curve, and $m_{y}^{BF}(I)$ at $A=0.4$ and $h_{R}=0$. All panels are done at $r=0.4$, $\omega=0.485$. The green hollow arrows indicate the value of current at witch the time dependence are calculated.  (b) Temporal dependence and FFT for  $V(t)$ and $m_{y}(t)$  component at $I=0.47$ and $I=0.491$ for increasing current direction (see hollow arrows in (a)) . }	
	\label{A1}
\end{figure}

In this section, we demonstrate additional analysis of the bifurcation and temporal dependence for the Shapiro and Buzdin steps separately with same simulation parameters shown in Fig.4 in the manuscript. As we see in Fig.\ref{A1}(a), the bifurcation along the Shapiro step shows disconnected branches (jumps) for $m_{y}^{BF}(I)$ around $I=0.5$ for decreasing current direction and around $I=0.475$ for increasing current direction. The question now, if this jumps on $m_{y}^{BF}(I)$ correspond to the two states of $m_{y}^{\pm}$ or not?. The answer to this question is shown in Fig.\ref{A1}(b), which presents the temporal dependence for $V(t)$ and $m_{y}(t)$ component at two current values before ($I=0.47$) and after ($I=0.491$) the jump for increasing current direction (the same for deceasing current, not shown here). As it is illustrated, the temporal dependence for $V(t)$ and $m_{y}(t)$ is uniform and the Fast Fourier transform, demonstrates frequency locking along the Shapiro step with frequency $V\equiv\omega_J=\omega_{R}=0.485$. We note that the amplitude of $V(t)$ and $m_{y}(t)$ as the current changes has a very small modulations along the Shapiro step, However the voltage average is constant and $m_{y}^{av}(I)$ is almost zero. This modulations is reflected on the values of the Poincar\'e section along the Shapiro step.

\begin{figure}[h!]
	\centering
	\includegraphics[width=0.45\linewidth]{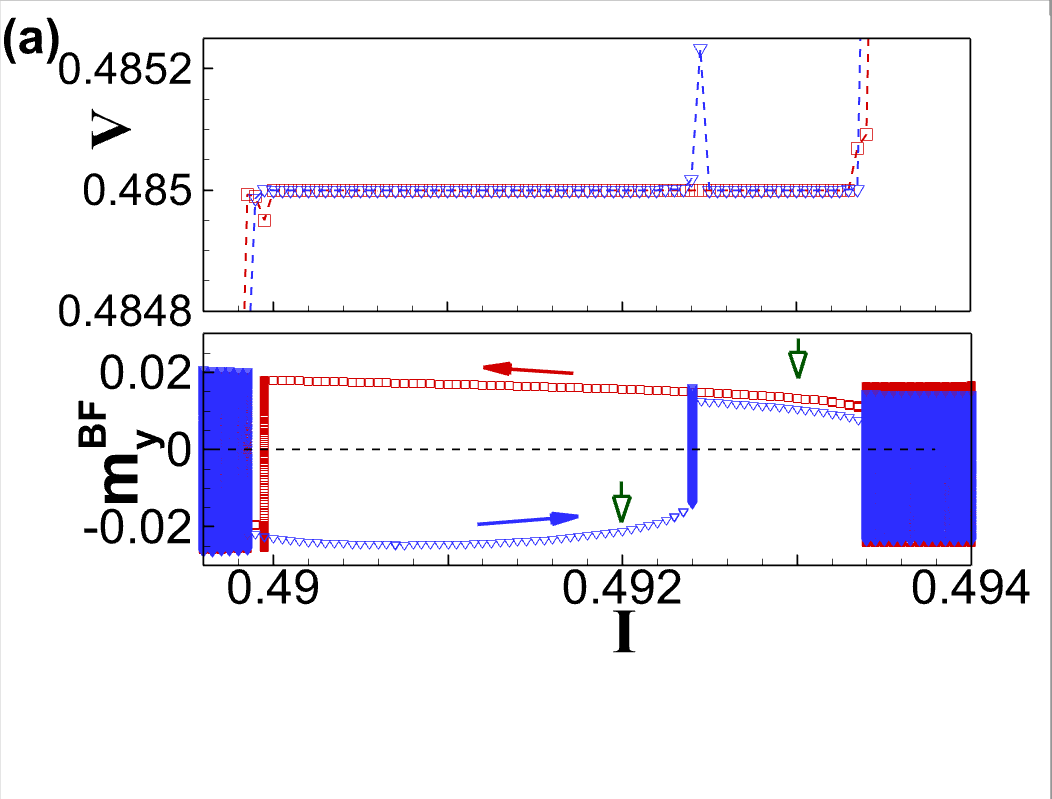}
	\includegraphics[width=0.44\linewidth]{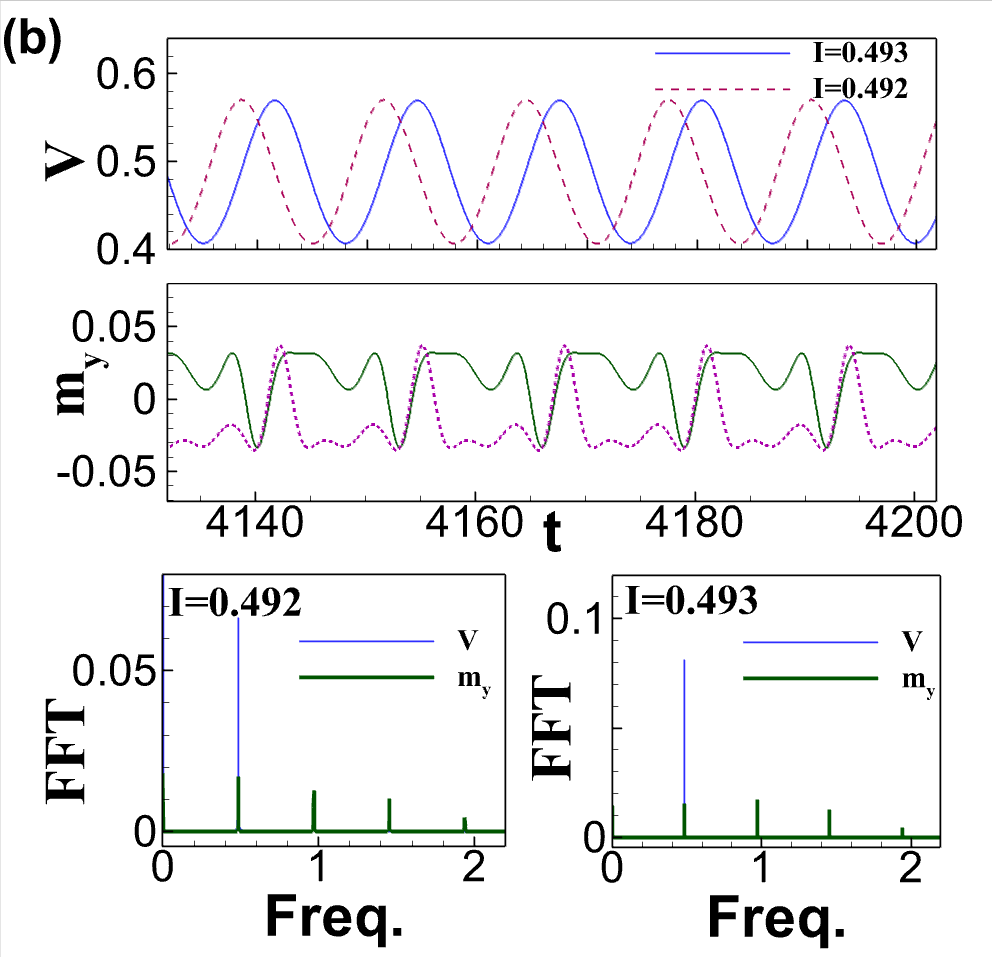}
	\caption{ (a) Enlarged part for Buzdin step with two loop calculation for $IV$-curve, and $m_{y}^{BF}(I)$ at $A=0$ and $h_{R}=1$. All panels are done at $r=0.4$, $\omega=0.485$. The green hollow arrows indicate the value of current at witch the time dependence are calculated. (b) Temporal dependence and FFT for $V(t)$ and $m_{y}(t)$ component at $I=0.492$ and $I=0.493$ for increasing current direction (see hollow arrows in (a)). }	
	\label{A2}
\end{figure}

Similar to chimera, the Buzdin step (see Fig.\ref{A2} ) demonstrates two bubble structure for $m_{y}^{max}(I)$ with the decreasing and increasing current. In difference with Shapiro step, the Poincar\'e section at the jump between the bifurcation branches demonstrate several points (higher order periodicity). Now, does this two branches at $I=0.4899$ and $I=0.4924$ corresponds to two states of $m_{y}$?. The answer to this question is illustrated in Fig.\ref{A2}(b), which presents the temporal dependence for $V(t)$ and $m_{y}(t)$ component at two current values before ($I=0.492$) and after ($I=0.493$) the jump at $I=0.4924$ for increasing current direction. As we see, the temporal dependence for $V(t)$ is uniform and the Fast Fourier transform demonstrate one frequency line at $V\equiv\omega_J=\omega_{R}=0.485$. However, $m_{y}$ temporal dependence shows oscillations with dieffrent amplitudes, and reflects the two states of $m_{y}^{max}(I)$ as in chimera step. Also, the FFT for $m_{y}(t)$ shows several frequency lines with dieffrent amplitudes at $\omega_J=n \omega_{R}$ and $n=1,2,..$ is integer. The same was observed at $I=0.4899$ (not shown here).

\subsection{Effect of pulse parameter}
\begin{figure}[h!]
	\centering
	\includegraphics[width=0.8\linewidth]{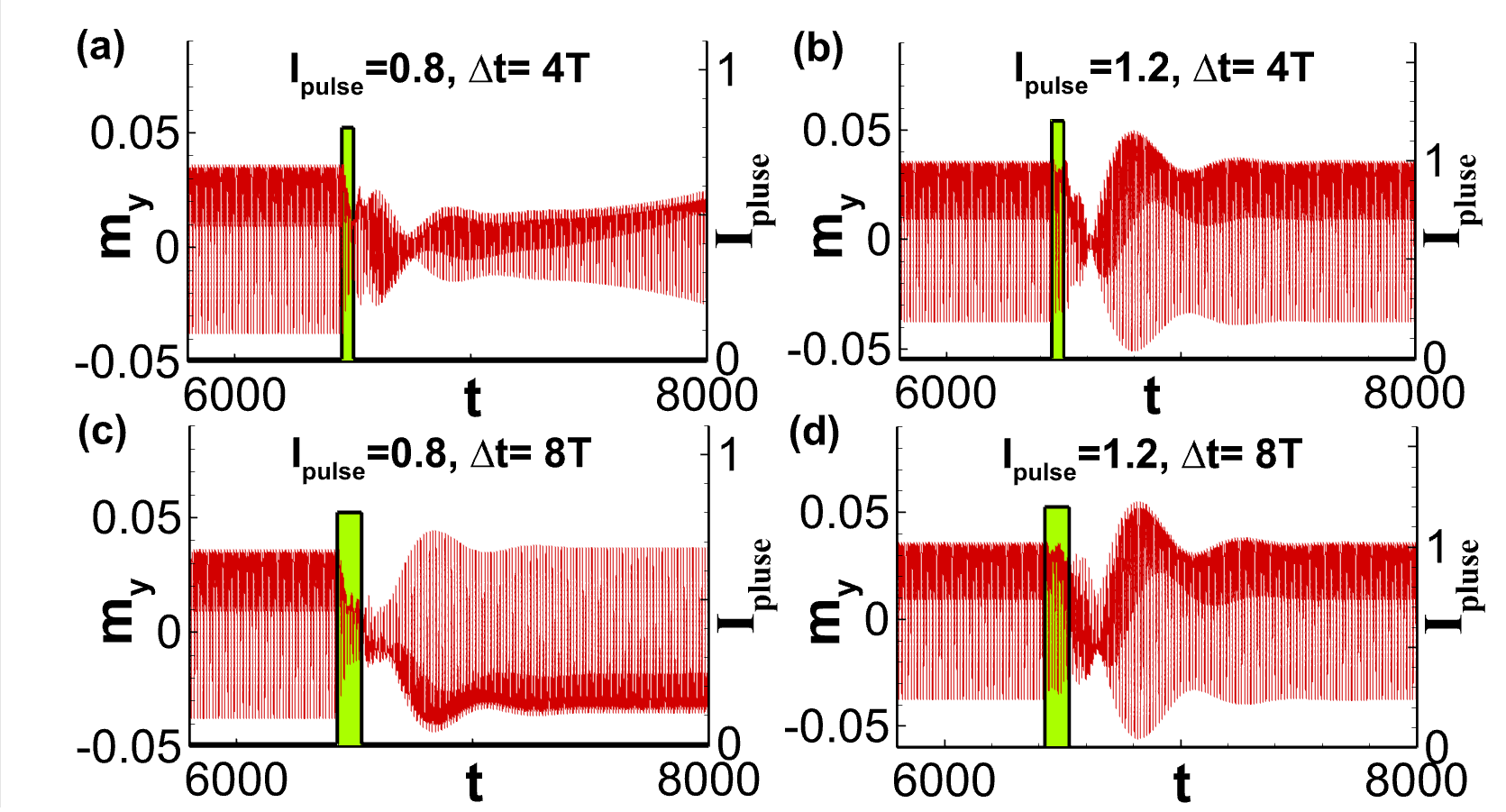}
	\caption{Effect of current pulse parameter on the switching between $m_{y}^{\pm}$ states. All panels are done at $r=0.4$, $G=0.01$, $\alpha=0.01$, $\omega_{R}=0.485$, and $\omega_{F}=0.5$.}
	\label{pulse_de}
\end{figure}

Fig.\ref{pulse_de} illustrate that one can control the switching between the states of $m_{y}$ by changing pulse height ($I_{pulse}$) or duration of the pulse. As can be seen in Fig.\ref{pulse_de} (a,b) for pulse duration $\Delta t=4T$, where $T$ is the periodic time of teh external microwave field, no witching was observed for $I_{pulse}=0.8$ and $1.2$. However, for $\Delta t=4T$ the switching occurs for $I_{pulse}=1.2$  (Fig.\ref{pulse_de} (c,d)).

\subsection{Phase shift along one bubble}

\begin{figure}[h!]
	\centering
	\includegraphics[width=0.5\linewidth]{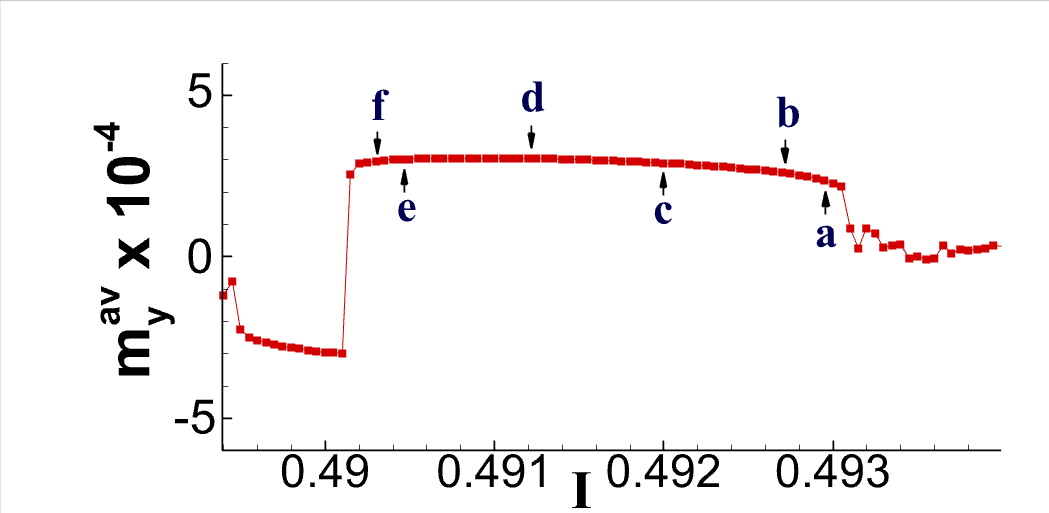}
	\caption{An enlarged part of $m_{y}^{av}(I)$ on Chimera step for decreasing current direction. Point "a" corresponds to $I=0.493$, "b" corresponds to $I=0.4925$, "c" corresponds to $I=0.492$, "d" corresponds to $I=0.4915$, "e" corresponds to $I=0.4908$, and "f" corresponds to $I=0.4905$.}
	\label{supp2}
\end{figure}

In this section, we calculate the phase shift and Pearson correlation between two points which lie on  same bubble for decreasing the current direction (see Fig.\ref{supp2}).  As demonstrated on Table.\ref{table2}, the phase shift grows from $14^{o}$ for the pair (a:b) to  $44^{o}$ for the pair (a:f). And the Pearson correlation (PC) has positive values ($0<PC<1$)  for all points on the same bubble.

\begin{table}[h!]
	\caption{\label{table2} Measurements of the Phase shift (PS) in degree and Pearsons correlation (PC) for several points shown in Fig.\ref{supp2}.}
	\begin{tabular}{|c|c|c|}
		\hline
		Pair & Phase Shift in $I_{s} (t)$& PC \\
		\hline
		ab & 14$^{o}$ & 0.965 \\
		\hline
		ac & 22$^{o}$ & 0.912 \\
		\hline
		ad & 29$^{o}$ & 0.844 \\
		\hline
		ae & 39$^{o}$ &0.708 \\
		\hline
		af &44$^{o}$ & 0.620 \\
		\hline
	\end{tabular}
\end{table}

The results demonstrated in Table.\ref{table2} show that, along one state of $m_{y}$, the phase shift increases and the  Pearsons correlation (PC) takes intermediate value. The same can take place by sweeping the current along the Shapiro step for the trivial Josephson junction with insulating barrier.

\end{document}